\documentclass[1p,preprint]{elsarticle}
\usepackage{amsmath}
\usepackage{amssymb}
\usepackage{amsmath}
\usepackage{color}
\usepackage{hyperref}
\newcommand\Let{\mathrel{\mathop:\!\!=}}
\newcommand{\iom}{{\ensuremath{i\omega}}}                                                                                                                                          
\def \sgn {\mathop {\rm sgn}}
\journal{Computer Physics Communications}

\begin{document}

\begin{frontmatter}

\title{Efficient implementation of the continuous-time hybridization expansion quantum impurity solver}

\author[X]{Hartmut Hafermann} \ead{hartmut.hafermann@cpht.polytechnique.fr}
\author[fribourg]{Philipp Werner} \ead{philipp.werner@unifr.ch}
\author[michigan,pks]{Emanuel Gull} \ead{egull@umich.edu}
%\address[eth]{ETH Zurich, 8093 Zurich, Switzerland}
\address[X]{\'{E}cole Polytechnique, CNRS, 91128 Palaiseau Cedex, France}
\address[fribourg]{Department of Physics, University of Fribourg, 1700 Fribourg, Switzerland}
\address[michigan]{University of Michigan, Ann Arbor, MI 48109, USA}
\address[pks]{Max Planck Institut f\"{u}r die Physik komplexer Systeme, 01187 Dresden, Germany}

\begin{abstract}
Strongly correlated quantum impurity problems appear in a wide variety of contexts ranging from nanoscience and surface physics to material science and the theory of strongly correlated lattice models, where they appear as auxiliary systems within dynamical mean-field theory. Accurate and unbiased solutions must usually be obtained numerically, and continuous-time quantum Monte Carlo algorithms, a family of algorithms based on the stochastic sampling of partition function expansions, perform well for such systems.
With the present paper we provide an efficient and generic implementation of the hybridization expansion quantum impurity solver, based on the segment representation.
We provide a complete implementation featuring most of the recently developed extensions and optimizations. 
Our implementation allows one to treat retarded interactions and provides generalized measurement routines based on improved estimators for the self-energy and for vertex functions. 
The solver is embedded in the ALPS-DMFT application package.
\end{abstract}

\end{frontmatter}
{\bf PROGRAM SUMMARY}

\begin{small}
\noindent
%% for information required in the program summary section, see
%% http://cpc.cs.qub.ac.uk/FAQanswers.html#Q4
%{\em Manuscript Title:} Hybridization Expansion CT-QMC solver \\ 
%{\em Authors:} Hartmut Hafermann, Philipp Werner, Emanuel Gull\\ 
{\em Program Title:} ct-hyb \\
{\em Catalogue identifier}: \verb#AEOL_v1_0#   \\
{\em Program summary URL:}\\
\verb#http://cpc.cs.qub.ac.uk/summaries/AEOL_v1_0.html#\\
{\em Program obtainable from:} CPC Program Library, QueenÕs University, Belfast, N. Ireland\\
{\em Licensing provisions:} Use of the hybridization expansion impurity solver requires citation of this paper. Use of any ALPS program requires citation of the ALPS \cite{ALPS20} paper.\\
%{\em No. of lines in distributed program, including test data, etc.:} 650044\\
%{\em No. of bytes in distributed program, including test data, etc.:} 20553265\\
{\em Distribution format:} tar.gz\\
{\em Programming language:} \verb*#C++#/\verb*#Python#\\
{\em Computer:} Desktop PC, high-performance computers.\\
{\em Operating system:} Unix, Linux, OSX, Windows.\\
{\em Has the code been parallelized?:} Yes, MPI parallelized.\\
{\em RAM:} 1 GB.\\
{\em Number of processors used:} 1 -- 1024\\ 
{\em Running time:} 1--8\, h\\
{\em Classification:} 7.3 \\
{\em Keywords:} DMFT, CT-QMC, CT-HYB.\\ 
{\em External routines/libraries:}  ALPS \cite{ALPS20,ALPS05,ALPS07}, BLAS \cite{BLAS79,BLAS02}, LAPACK \cite{LAPACK99}, HDF5  \cite{hdf5} .\\ 
{\em Nature of problem:}
Quantum impurity models were originally introduced to describe a magnetic transition metal ion in a non-magnetic host metal. They are widely used today. In nanoscience they serve as representations of quantum dots and molecular conductors. In condensed matter physics, they are playing an increasingly important role in the description of strongly correlated electron materials, where the complicated many-body problem is mapped onto an auxiliary quantum impurity model in the context of dynamical mean-field theory, and its cluster and diagrammatic extensions. They still constitutes a non-trivial many-body problem, which takes into account the (possibly retarded) interaction between electrons occupying the impurity site. Electrons are allowed to dynamically hop on and off the impurity site, which is described by a time-dependent hybridization function.\\
{\em Solution method:}
The quantum impurity model is solved using a continuous-time quantum Monte Carlo algorithm which is based on a perturbation expansion of the partition function in the impurity-bath hybridization. Monte Carlo configurations are represented as segments on the imaginary time interval and individual terms correspond to Feynman diagrams which are stochastically sampled to all orders using a Metropolis algorithm. For a detailed review on the method, we refer the reader to [8].
\end{small}

\section{Introduction}

Quantum impurity models describe a set of correlated sites or orbitals embedded in a bath of non-interacting states.
Quantum impurity models appear in a range of contexts, including magnetic impurities embedded in a non-magnetic host material \cite{Anderson61}, nanoscience, where they are used to describe quantum dots and molecular conductors \cite{Hanson07}, and surface science \cite{Brako81}, for the description of molecules adsorbed on a substrate.
Quantum impurity solvers are also an essential ingredient of the dynamical mean -field (DMFT) \cite{Metzner89,Georges92,Georges96,Georges04} approximation to correlated lattice systems, which has had enormous success in recent years in the simulation of correlated material systems \cite{Held06,Held07,Kotliar06} and lattice models \cite{Maier05}.

With this article we provide a description and a state-of-the-art implementation of the continuous-time \cite{Rubtsov04,Rubtsov05,Gull11_review} `hybridization expansion' quantum Monte Carlo impurity solver for density-density interactions \cite{Werner06}. Our implementation includes in particular the important numerical and conceptual advances developed over the last few years: improved estimators \cite{Hafermann12}, frequency and Legendre measurements \cite{Boehnke11}, measurement of vertex functions \cite{Gull11_review}, treatment of retarded interactions \cite{Werner07holstein,Werner10_frequency}, and parallelization to a large number of cores \cite{ALPS20,ALPS05,ALPS07,Gull10_submatrix}.

General fermionic impurity models have the form $H_\text{imp}=H_\text{loc}+H_\text{bath}+H_\text{hyb}$, where
\begin{align}
H_\text{loc}&=\sum_{ab} E^{ab}d^\dagger_a d_b+\sum_{abcd}U^{abcd}d^\dagger_a d^\dagger_b d_c d_d,\label{Hloc}\\
H_\text{bath}&=\sum_{k\alpha} \varepsilon_{k\alpha}c^\dagger_{k\alpha}c_{k\alpha},\label{Hbath}\\
H_\text{hyb}&=\sum_{k\alpha b}{V}_k^{\alpha b}c^\dagger_{k\alpha}d_b +\text{h.c.}. \label{Hhyb}
\end{align}
The term $H_\text{loc}$ corresponds to the impurity with level energies and intra-orbital hoppings described by $E$ and the interaction terms parametrized by $U$ (roman indices label the different interacting orbitals including spin). $H_\text{bath}$ describes the non-interacting bath with quantum numbers $k$ and spin/orbital index $\alpha$. The hybridization term $H_\text{hyb}$ represents the exchange of electrons between the impurity and the bath, parametrized by the hybridization matrix $V_k^{\alpha b}$. All the relevant properties of the bath are encoded in the hybridization functions
\begin{equation}
\Delta_{ab}(\iom_n)=\sum_{k,\alpha}\frac{{V_k^{a\alpha}}^*V_k^{\alpha b}}{\iom_n-\varepsilon_{k\alpha}}.
\end{equation}
In a DMFT calculation, these parameters are determined self-consistently.

The observable of primary interest is the impurity Green's function $G$, defined as\footnote{We adopt here the `many-body' definition of $G$ with $G(0_+)<0$ as opposed to the `Monte Carlo' definition where $G(0_+)>0$.}
\begin{equation}
G_{ab}(\tau-\tau') \Let -\langle T_{\tau} c_{a}(\tau)c_{b}^\dagger(\tau')\rangle .
\label{gtau}
\end{equation}
The Green's function contains the information about all single-particle properties of the model, including the spectral function and the particle density.  Higher-order correlation functions are needed for the measurement of susceptibilities and the calculation of the (reducible or irreducible) vertex. 

\section{Hybridization Expansion}
In recent years, a new class of efficient Monte Carlo techniques has been developed for solving quantum impurity models: the continuous-time impurity solvers \cite{Gull11_review,Rubtsov04,Rubtsov05,Werner06,Werner06Kondo,Haule07,Gull08_ctaux}. The hybridization expansion \cite{Werner06,Werner06Kondo} approach that we describe here is particularly well suited for the single- and multi-orbital impurity problems that typically appear in single-site DMFT calculations. In this approach, the partition function $Z=\text{Tr}_{d,c} [e^{-\beta H_\text{imp}}]$ is expanded in powers of the hybridization term $H_\text{hyb}$. Monte Carlo configurations consist of a sequence of creation and annihilation operators, and thus represent a sequence of hybridization events (electrons hopping from the bath to the impurity or back into the bath). The Monte Carlo weight $w(\{\tau_i\})$ of such a configuration is the product of two factors, 
\begin{equation}
w(\{\tau_i\})=w_\text{loc}(\{\tau_i\})w_\text{hyb}(\{\tau_i\}),
\end{equation}
where $w_\text{loc}(\{\tau_i\})$ is the trace over the impurity states for the specific sequence of impurity creation and annihilation operators, and  $w_\text{hyb}(\{\tau_i\})$ is the trace over the bath states. Because the bath is non-interacting, the latter can be expressed as the determinant of a matrix $M^{-1}$, whose elements are hybridization functions connecting the different pairs of creation and annihilation operators. 

The local contribution is evaluated explicitly. For a generic model, i.~e. if the Hamiltonian is not diagonal in the occupation number basis, this is a computationally expensive procedure which scales exponentially with system size \cite{Werner06Kondo,Haule07, Laeuchli09}, even though substantial progress has been made to make these simulations affordable \cite{Haule07,Huang12b,Gull11_review,Parragh12}. An important simplification occurs if the interaction term is restricted to density-density interactions. In this case, the atomic part of the Hamiltonian takes the form
\begin{equation}
H_\text{loc}=\sum_{a} E^{a}n_a + \sum_{ab}U^{ab}n_a n_b
\label{Hlocdens}
\end{equation}
and the local trace can be evaluated in polynomial time \cite{Werner06}.

\section{Implementation}
We aim to provide a fast but general implementation of this algorithm. We include a formalism for retarded interactions, efficient measurements of single-particle Green's functions in imaginary time in both the Matsubara and Legendre basis, measurements of two-particle Green's functions, improved estimators for the self-energy and vertex functions, and energy and sector statistics measurements, as well as density-density correlation functions. 
We briefly describe each of these features in the following. More detailed explanations can be found in a recent review \cite{Gull11_review} and the original papers.

\subsection{Retarded Interactions}

Retarded interactions appear in models with electron-phonon coupling (e.g. the Hubbard-Holstein model \cite{Holstein59}) or in realistic material simulations with dynamically screened Coulomb interaction. They also arise in the context of extended dynamical mean-field theory \cite{Smith00,Chitra01} and its diagrammatic extension, the dual boson approach \cite{Rubtsov12}.
An efficient technique to treat retardation effects in the hybridization expansion  has been described in Refs.~\cite{Werner07holstein} and \cite{Werner10_frequency}. In order to treat a model with a frequency-dependent local Coulomb interaction $U(\omega)$ one has to (i) set the instantaneous interactions 
to the screened value $U(\omega=0)$ and (ii) introduce an effective retarded interaction $K(\tau)$ between all pairs of creation and annihilation operators (irrespective of flavor). The explicit expression for the function $K(\tau)$ in the interval $0\le\tau\le\beta$ is \cite{Werner10_frequency}
\begin{align}
K(\tau)&=\int_0^\infty \frac{d\omega}{\pi}\frac{\text{Im}U(\omega)}{\omega^2}[W_{\omega}(\tau)-W_{\omega}(0)],
\label{Wdef}
\end{align}
with $W_{\omega}(\tau)=\cosh\big[\big(\tau-\frac{\beta}{2}\big)\omega\big]/\sinh\big[\frac{\omega\beta}{2}\big]$. The sign of the retarded interaction depends on the type of operators (creators/annihilators) which are connected, so the additional weight factor of the Monte Carlo configuration becomes
\begin{equation}
w_\text{screen}(\{\tau_i\})=e^{\sum_{2n\geq i>j\geq1}s_is_jK(\tau_i-\tau_j)},
\end{equation}
where the sum is over all the operator pairs $i,j$ and $s_i=+1$ if operator $i$ is a creation operator and $-1$ if it is an annihilation operator.

Frequency-dependent (retarded) interactions can thus be incorporated very easily into the hybridization expansion method. If a new pair of creation and annihilation operators (segment) is introduced into a Monte Carlo configuration, one computes the change in $w_\text{loc}$ (using $U(\omega=0)$) and the change in $w_\text{screen}$. For the latter, one has to evaluate the retarded interactions between all the operator pairs which involve at least one of the newly inserted creation and annihilation operators. The computational effort for a local update is O($n$) and therefore negligible compared to the evaluation of the determinant ratio, which is O($n^2$).

\subsection{Green's function measurements}

\subsubsection{Imaginary time measurement}

For a given inverse hybridization matrix $M$, the Green's function, Eq.~(\ref{gtau}), is measured on the interval $[0,\beta]$ as
\begin{equation}
G_{ab}(\tau) =\!\left\langle\frac{-1}{\beta}\sum_{\alpha\beta=1}^{n}\!\!M_{\beta\alpha}\delta^{-}(\tau,\tau_{\alpha}^{e}-\tau_{\beta}^{s})\delta_{a\alpha}\delta_{b\beta}\!\!\right\rangle
\label{gtaumeas}
\end{equation}
where $\langle\ldots\rangle$ denotes the Monte Carlo average and $\delta^{-}(\tau,\tau')\Let \sgn(\tau')\delta(\tau-\tau'-\theta(-\tau')\beta)$ accounts for the antiperiodicity of $G(\tau)$. 
The $\tau^e$ ($\tau^s$) denote the times of annihilation (creation) operators. 
The Green's function is measured on a discrete grid representing the continuous variable $\tau$. Since the complexity of the algorithm does not depend on the grid resolution, this grid can be chosen arbitrarily fine.

\subsubsection{Matsubara frequency measurement}

It is often convenient to use the Matsubara frequency representation of the Green's function. Carrying out the Fourier transformation of Eq.~(\ref{gtaumeas}) analytically, the Matsubara coefficients $G(i\omega_{n})$ of $G(\tau)$ are measured directly, without the need of a time-discretization grid:
\begin{equation}
G_{ab}(i\omega_{n}) =\!\!\left\langle\frac{-1}{\beta}\sum_{\alpha\beta=1}^{n}\!\!M_{\beta\alpha}e^{\iom_{n}(\tau_{\alpha}^{e}-\tau_{\beta}^{s})}\delta_{a\alpha}\delta_{b\beta}\!\!\right\rangle.
\label{gwmeas}
\end{equation}
%This is superseded by nfft, better not to describe too much of an obsolete technology
%
%We use a fast update of the exponentials where the exponentials for all frequencies are computed that, for a given time difference, the \emph{exp()} function is explicitly evaluated only once.
%[??? HOW EXACTLY DOES THIS WORK ???]

\subsubsection{Orthogonal polynomial representation}
A measurement procedure based on an expansion of the Green's function on the interval [0,$\beta$] in terms of orthogonal polynomials was introduced in Ref.~\cite{Boehnke11}. 
We focus here on Legendre polynomials. Again, the transformation of the measurement rule Eq.~\eqref{gtaumeas} is carried out analytically. 
The Legendre transformation is \emph{unitary} and can be written in terms of a matrix multiplication $G(i\omega_{n}) = \sum_{l\geq0} T_{nl} G_{l}$, with
expansion coefficients
\begin{equation}
G_{ab;l} = \left\langle\frac{-\sqrt{2l+1}}{\beta}\!\!\sum_{\alpha\beta=1}^{n}\!\!M_{\beta\alpha}\tilde{P}_{l}(\tau_{\alpha}^{e}-\tau_{\beta}^{s})\delta_{a\alpha}\delta_{b\beta}\right\rangle
\label{glmeas}
\end{equation}
where $\tilde{P}_{l}(\tau)\Let P_{l}[x(\tau)]$ for $\tau>0$ and $-P_{l}[x(\tau+\beta)]$ for $\tau<0$ and $x(\tau)=2\tau/\beta-1$ maps the interval $[0,\beta]$ to $[-1,1]$. 
Observable estimates exhibit a clear plateau as a function of $l$, making a basis truncation well controlled. 
The coefficients $G_{l}$ generally decay faster than any power of $1/l$, leading to a highly compact representation of observables. 
As the noise is mostly carried by high-order coefficients, this representation  acts as an efficient noise filter. 

\subsection{Improved estimators}
An important physical quantity is the impurity model self-energy. It is usually computed by inverting the interacting and non-interacting Green's functions of the impurity according to the Dyson equation
\begin{equation}
\Sigma(\iom_{n}) = G_{0}^{-1}(\iom_{n}) - G^{-1}(\iom_{n}) .
\end{equation}
This inversion amplifies statistical noise, in particular at high frequencies where the difference between the Green's functions $G$ and $G_0$ (which both decay as $1/\omega_{n}$ with constant errors) is small.

This problem has been resolved recently \cite{Hafermann12}. Using the equation of motion, the \emph{product} of the self-energy and Green's function can be expressed in terms of an additional correlator $F^{j}(\tau-\tau')\Let-\langle T_{\tau}c_{a}(\tau)c^{\dagger}_{b}(\tau')n_{j}(\tau')\rangle$ (so-called ``Bulla's trick'' \cite{Bulla98}):
\begin{equation} 
(G\Sigma)_{ab}(\iom_{n}) = \frac{1}{2}\sum_{j}(U_{jb} + U_{bj}) F^{j}_{ab}(\iom_{n}).
\label{gsigma}
\end{equation}
The additional quantity can be computed from $G$ at minimal extra cost, both in the Matsubara and Legendre basis (cf. Eqs. \eqref{gwmeas},\eqref{glmeas}). Combining the Legendre filter with the improved estimator yields self-energies with unprecedented accuracy (see Sec. \ref{sec:examples} for an illustration).
The current implementation has been generalized to evaluate the improved estimators in the presence of retarded interactions. Details will be published elsewhere.

\subsection{Two-particle Green's functions}

Two-particle Green's functions and the related vertex function have gained importance in recent years, partly due to an increase in computational resources but also through the development of new methods. They serve as input for the calculation of susceptibilities within DMFT and provide the basis for novel extensions of DMFT \cite{Toschi07,Rubtsov08,Hafermann09}.
To meet these requirements, we provide measurements for the two-particle Green's function
\begin{align}
G^{(2)}&(\tau_{a},\tau_{b},\tau_{c},\tau_{d})\nonumber\Let\langle T_{\tau} c_{a}(\tau_{a})c^{\dagger}_{b}(\tau_{b})c_{c}(\tau_{c})c^{\dagger}_{d}(\tau_{d})\rangle .
\end{align}

These functions are measured in frequency space, as a function of two fermionic frequencies and one bosonic frequency. We further provide the measurement of the correlation function
\begin{align}
H^{j}&(\tau_{a},\tau_{b},\tau_{c},\tau_{d})\Let\nonumber\\&\langle T_{\tau} n_{j}(\tau_{a}) c_{a}(\tau_{a})c^{\dagger}_{b}(\tau_{b})c_{c}(\tau_{c})c^{\dagger}_{d}(\tau_{d})\rangle .
\end{align}
which allows one to accurately compute the vertex function from an improved estimator expression in analogy to Eq.~(\ref{gsigma}). For details, see Ref.~\cite{Hafermann12}.

\subsection{LDA+DMFT interface}
The LDA+DMFT method \cite{Kotliar06,Held07} provides an interface between band structure methods and many-body theory: A local density approximation (LDA) band structure is obtained from a band theory calculation, to which two-body correlations (typically static Coulomb repulsion and Hunds coupling terms) are added. Our code has been designed specifically with these applications in mind, and it provides an interface to specify interaction matrices and double-counting corrections. In the context of LDA+DMFT, off-diagonal hybridizations and non-density-density interactions may become important. While related algorithms exist to treat them \cite{Werner06Kondo, Laeuchli09}, and substantial progress has been made to make these simulations affordable \cite{Haule07,Huang12b,Gull11_review,Parragh12}, the present implementation is restricted to diagonal hybridization functions and density-density interactions.

\subsection{Sector statistics}
A histogram of the atomic states occupied in the course of the simulation may provide important physical insights \cite{Haule07}. An impurity with a single orbital can assume one of four different states: empty ($|0\rangle$), singly occupied ($|\!\!\uparrow\rangle$ or $|\!\!\downarrow\rangle$) or completely filled ($|\!\uparrow\downarrow\rangle$). The algorithm measures the fraction of time the impurity spends in any given atomic state and collects a histogram. A typical example is shown in Fig. \ref{fig:sector_stats}.

\subsection{Parallelization}
Continuous-time Monte Carlo codes are straightforwardly parallelizable by running a different random walk on each core. 
The cost of thermalization and the non-parallel sections of the code are negligible, therefore efficient implementations scale up to a comparatively large number of CPUs \cite{Gull10_submatrix}. The present implementation of the hybridization expansion code supports MPI parallelization using the next-generation ALPS scheduler \cite{ALPS20}. Efficient use of collective communications and state-of-the-art binary data storage \cite{hdf5} mean that for typical applications the code scales to more than $1000$ cores without noticeable overhead.

\subsection{Python interface}
The solver can be built either as a standalone executable or as a Python module. Using the solver as a Python module allows one to perform all computationally expensive operations in C++,
while other aspects of the calculation (e.g. a DMFT self-consistency cycle, data evaluation, or application specific computations) can be written in Python. All data storage is implemented using the popular \verb#hdf5# library, with an interface both to C++ and Python.  

\section{Examples}
\begin{figure}[t]
\begin{center}
\includegraphics[scale=1,angle=0]{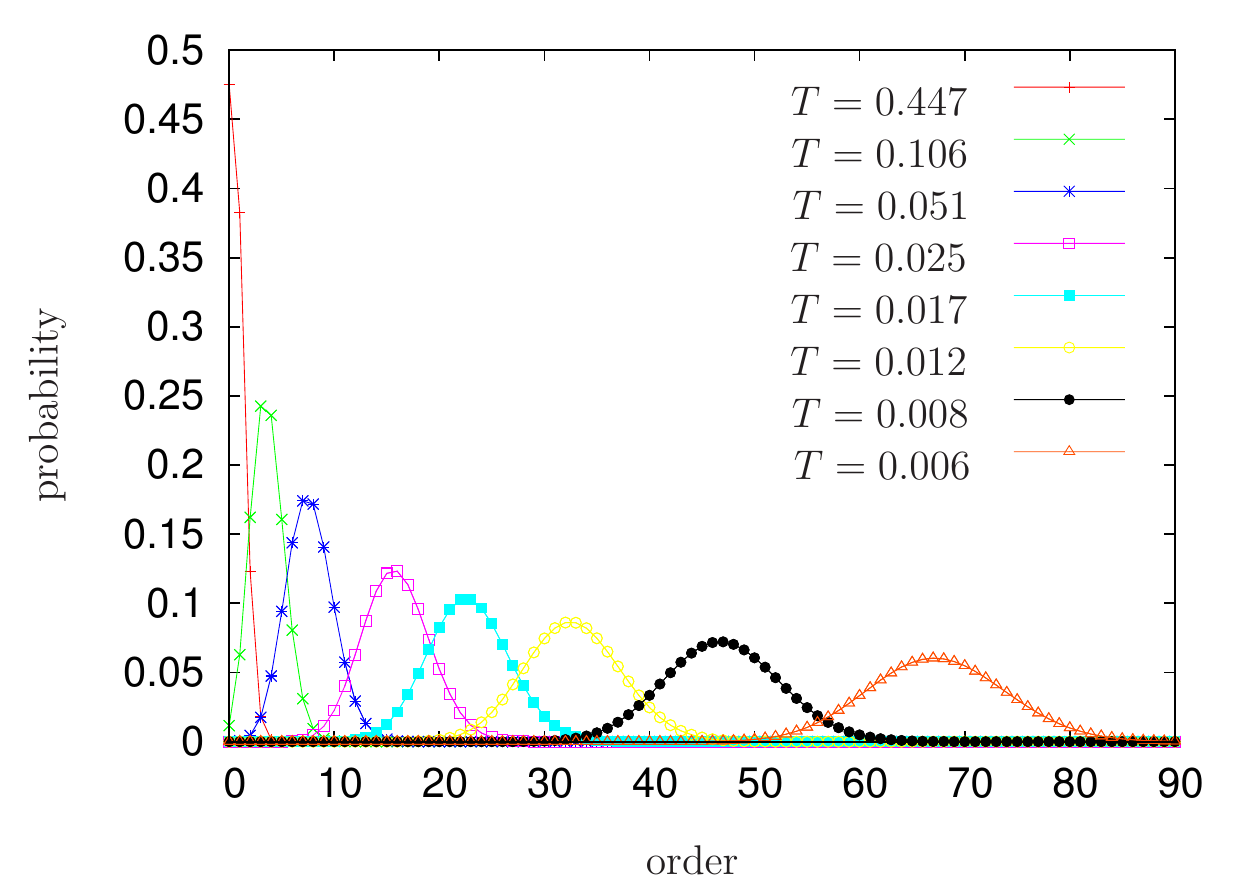} 
\end{center}
\caption{\label{fig:histogram} (Color online) Perturbation-order histograms for the single-orbital Anderson model with $U=2$ for different temperatures.
}
\end{figure}

\label{sec:examples}

\subsection{Single-orbital Anderson impurity model}

As a first example, we consider the single-orbital Anderson impurity model. In the following, we always use a semielliptical density of states with bandwidth equal to $4t$, and set $t=1$.
Fig. \ref{fig:histogram} shows the perturbation-order histograms for $U=2$ and different temperatures. The mean perturbation order shifts to larger values with decreasing temperature. Note that the average perturbation order yields the kinetic energy divided by temperature \cite{Haule07}. 

\begin{figure}[t]
\begin{center}
\includegraphics[scale=1,angle=0]{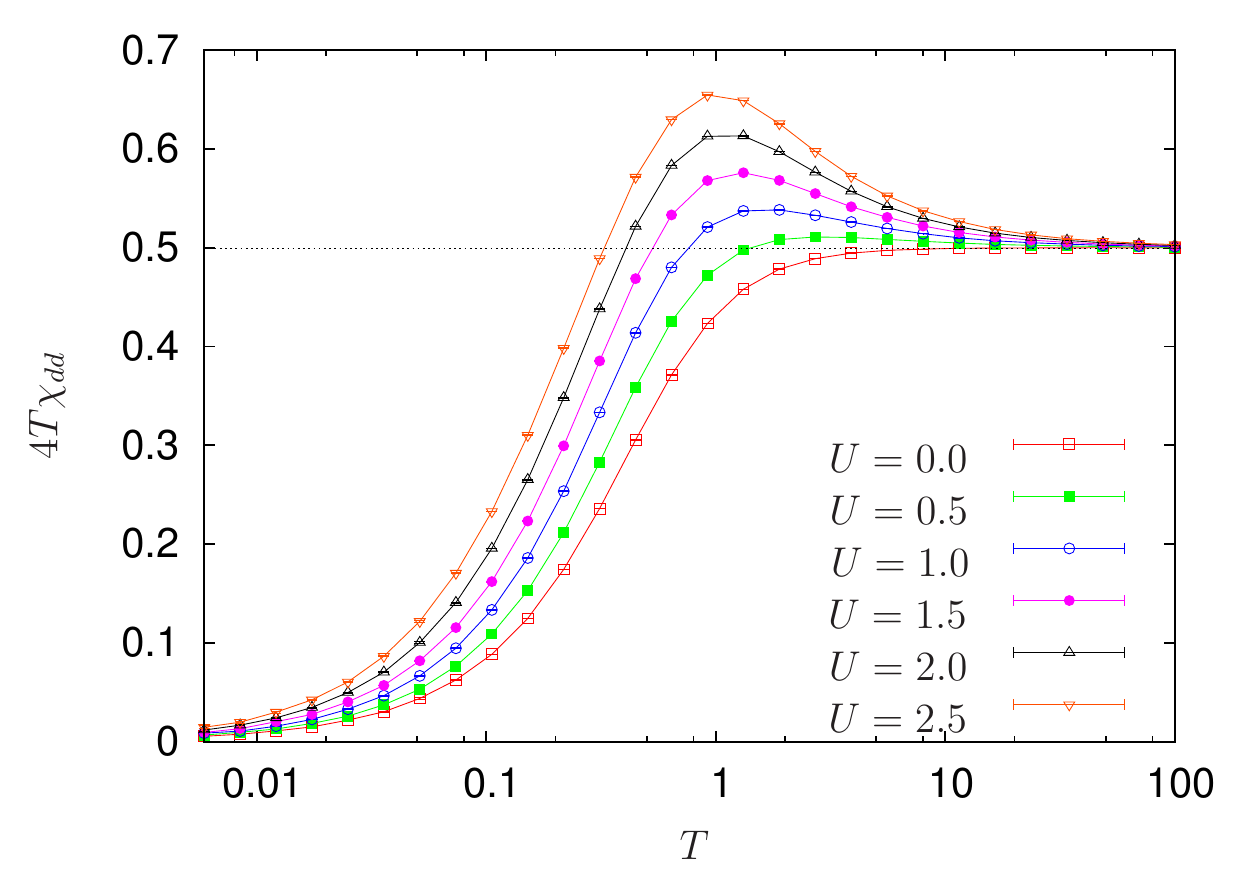} 
\end{center}
\caption{\label{fig:chiT} (Color online) Effective local moment of the single-orbital Anderson model as a function of temperature. As the temperature is lowered the local moment is formed and then decreases due to Kondo screening.
}
\end{figure}

As an example for a non-trivial observable, we measure the local spin-spin susceptibility 
\begin{equation}
\chi_{dd}(i\omega_{m})\Let \int_{0}^{\beta}d\tau \langle  S_{z}(\tau)S_{z}(0)\rangle e^{i\omega_{m}\tau}
\end{equation}
directly in frequency.
Fig \ref{fig:chiT} shows its static component $\chi_{dd}(\omega=0)$ times temperature (i.e., the effective local moment) versus temperature. As the temperature is lowered a local moment is formed, the magnitude of which increases with $U$. As the temperature is lowered further, the moment decreases as a result of screening due to the Kondo effect. The screening occurs on a scale set by the Kondo temperature $T_{K}$. With the present solver, it is possible to reach very low temperatures.

\subsection{Retarded interactions}

\begin{figure}[t]
\begin{center}
\includegraphics[scale=1,angle=0]{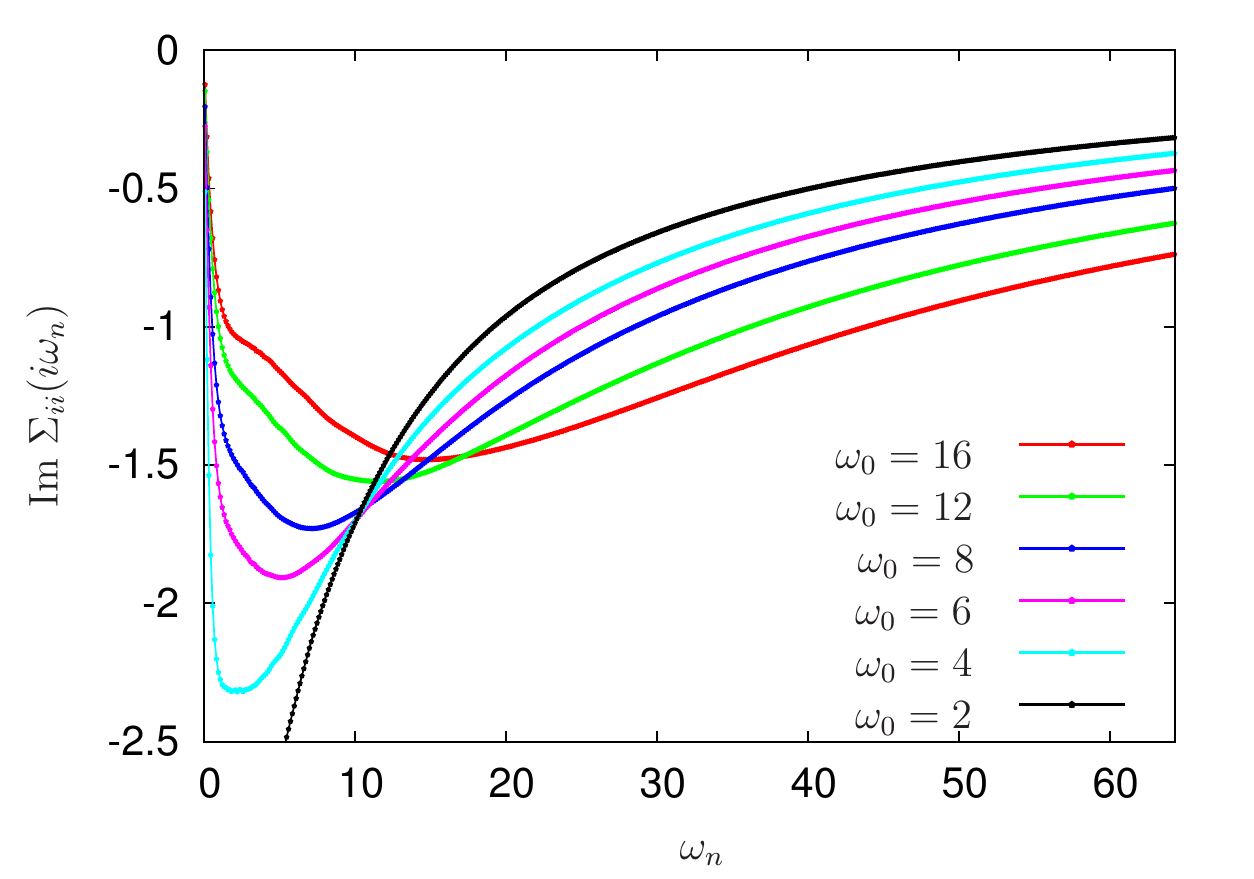} 
\end{center}
\caption{\label{fig:retint} (Color online) Self-energy of the Hubbard Holstein model with $U=8$, fixed screened interaction $U_{\text{scr}}=3$, temperature $T=0.02$ and different phonon frequencies $\omega_{0}$. The self-energy has been measured using improved estimators in the Legendre basis.
}
\end{figure}

As a second application, we consider a single-orbital model with a retarded interaction, corresponding to a Holstein-Hubbard model (``plasmon" frequency $\omega_0$, electron-boson coupling strength $\lambda$), which we solve within DMFT on the Bethe lattice.
The retarded interaction function $K(\tau)$ and its derivative are given by
\begin{align*}
 K(\tau) &= - \frac{\lambda^{2}}{\omega_{0}^{2}}\{\cosh[\omega_{0}(\beta/2-\tau)]/\sinh(\omega_{0}\beta/2)\} + c,\\
 K'(\tau) &= + \frac{\lambda^{2}}{\omega_{0}}\sinh[\omega_{0}(\beta/2-\tau)]/\sinh(\omega_{0}\beta/2),
\end{align*}
where the constant $c$ is chosen such that $K(0_{+})=0$.
Fig.~\ref{fig:retint} shows the electronic self-energy for fixed bare interaction $U=8t$ and fixed screened interaction $U_{\text{scr}}=U-2\lambda^{2}/\omega_{0}=3t$ for different screening frequencies $\omega_{0}$ at temperature $T=0.02t$. For large screening frequencies the system is metallic, while it undergoes a metal-insulator transition as $\omega_{0}$ is lowered. 
The corresponding results for the spectral function of this model are shown in Ref.~\cite{Werner10_frequency}.

\subsection{DMFT for Multi-Orbital Models}

\begin{figure}[t]
\begin{center}
\includegraphics[scale=1,angle=0]{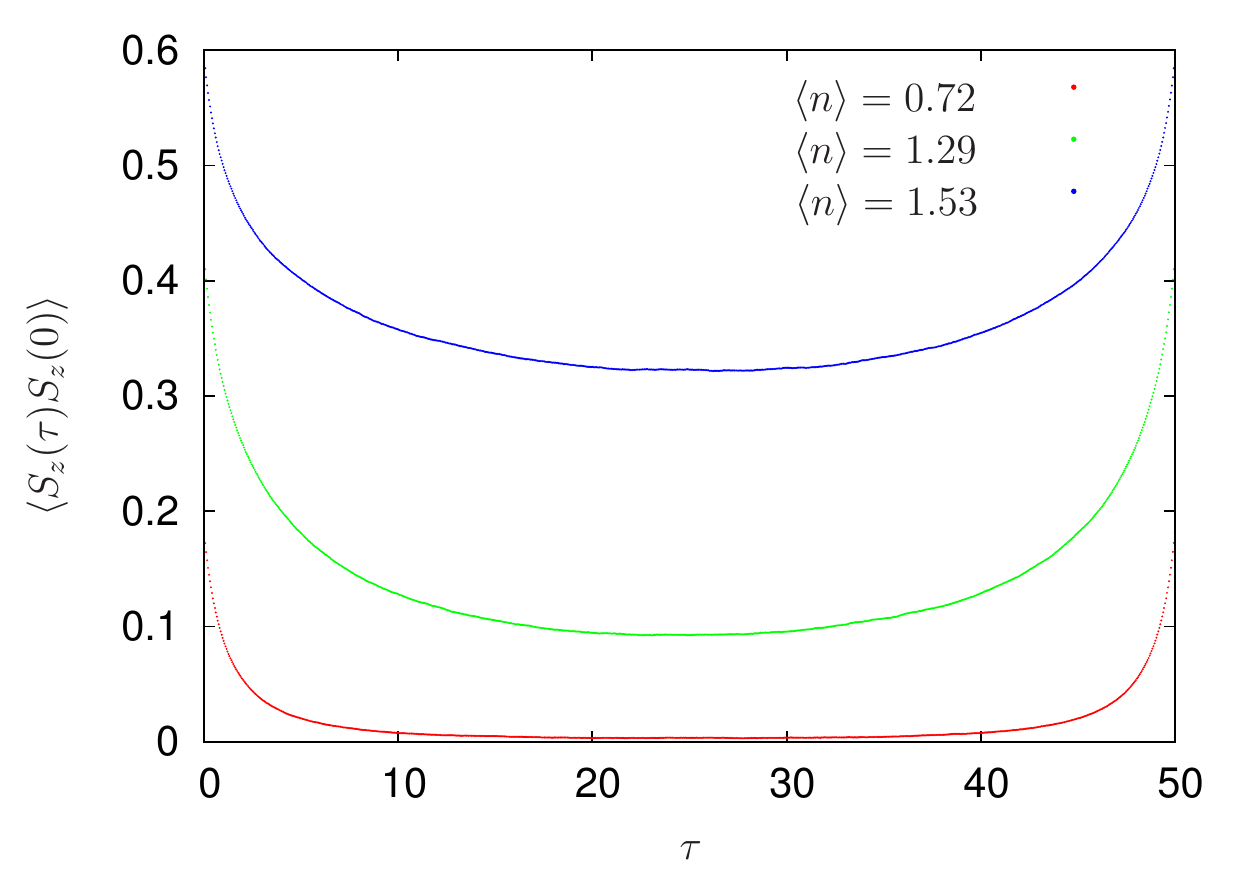} 
\end{center}
\caption{\label{fig:szszt} (Color online) Spin-spin correlation function as a function of imaginary time for various values of filling. Two-orbital model, $U/t=8$, $J/U=1/6$, $\beta t=50$.
}
\end{figure}

The code is well suited for studying multi-orbital problems in the density-density approximation. As an example we consider a two-orbital model with interaction 
\begin{align}
\sum_{ab}U^{ab}n_{a}n_{b} = &\ U\!\!\sum_{\alpha=1,2} n_{\alpha\uparrow}n_{\alpha\downarrow} + U'\sum_{\sigma}n_{1,\sigma}n_{2,-\sigma} \nonumber\\
& + (U'-J) \sum_{\sigma} n_{1,\sigma}n_{2,\sigma}.
\end{align}
Here $\sigma$ and $\alpha$ denote spin and orbital indices, $U$ and $U'=U-2J$ are the intra- and inter-orbital Coulomb interaction parameters and $J$ is the Hund's rule coupling coefficient.
We solve the model on the Bethe lattice with bandwidth $4t$ at temperature $T/t=1/50$, $U/t=8$ and relatively large Hund's coupling $J/U=1/6$. We consider only paramagnetic and paraorbital solutions. This model has been shown to exhibit a spin-freezing transition as a function of filling \cite{Hafermann12}.
This manifests itself for example in the spin-spin correlation function (see Fig. \ref{fig:szszt}), which is small for large times (i.e. near $\beta/2$) in the Fermi liquid phase, but approaches a large non-zero value in the frozen moment phase.

\begin{figure}[t]
\begin{center}
\includegraphics[scale=1,angle=0]{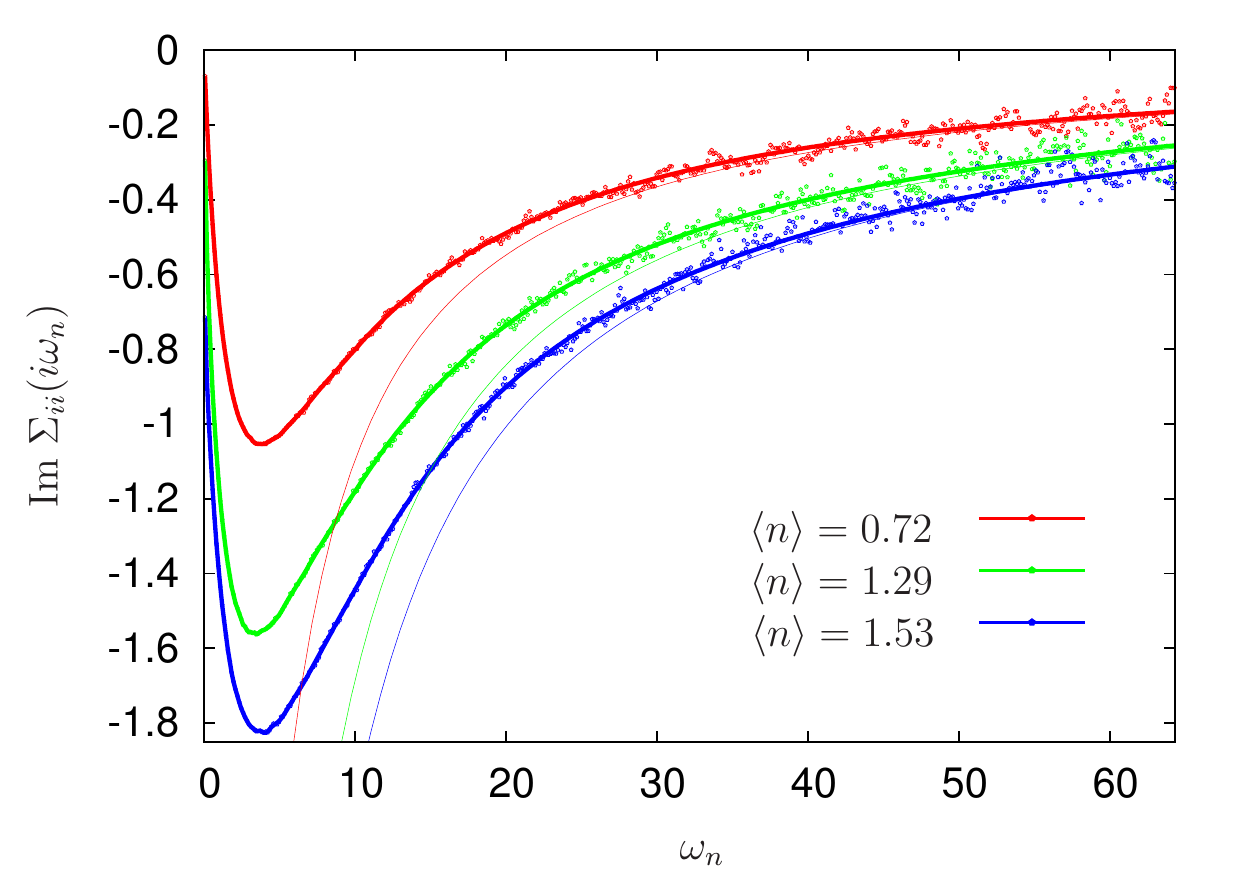}
\end{center}
\caption{\label{fig:sigma} (Color online) Self-energy of a two-orbital model ($U/t=8$, $J/U=1/6$, $T/t=0.02$) for different fillings, calculated using the improved estimators. Dots represent data from the Matsubara measurement, whereas the result from the Legendre measurement is shown by thick solid lines. Thin solid lines correspond to the high-frequency tails.
}
\end{figure}

As a second example we show the imaginary part of the self-energy for different fillings across the transition (Fig. \ref{fig:sigma}), determined using the improved estimator measured in two different bases: the Matsubara basis and the Legendre basis ($N_{l}=80$ coefficients).
The imaginary part of the high-frequency tails $\Sigma_\text{tail}(i\omega_{n})=\Sigma^{0}_{ii}+\Sigma^{1}_{ii}/(i\omega_{n})$ is shown for comparison. It has been calculated from the orbital densities $\langle n_{i}\rangle$ and equal-time density-density correlation functions $\langle n_{i}n_{j}\rangle$ using the expression
\begin{align}
\Sigma^{1}_{ii}=\sum_{kl}U_{ik}U_{il}(\langle n_{k}n_{l}\rangle - \langle n_{k}\rangle \langle n_{l}\rangle).
\end{align}
One can see that the Legendre representation efficiently filters the Monte Carlo noise and correctly reproduces the high-frequency tail.

\begin{figure}[t]
\begin{center}
\includegraphics[scale=1,angle=0]{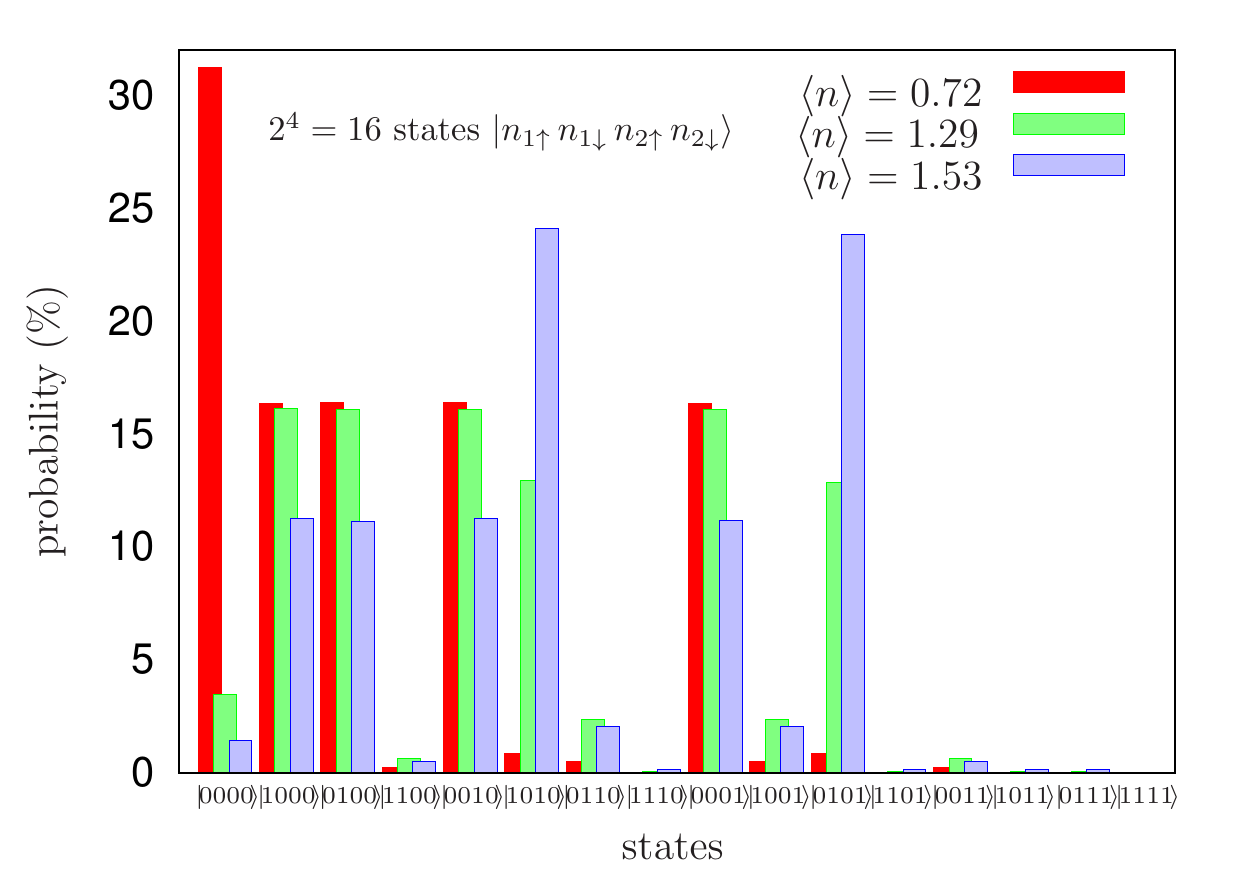} 
\end{center}
\caption{\label{fig:sector_stats} (Color online) Sector statistics for the two-orbital model ($U/t=8$, $J/U=1/6$, $T/t=0.02$) showing the probability of finding the impurity in a given atomic state. The states are labeled by the occupation numbers of the four distinct spin-orbital states.
}
\end{figure}

For the same parameters, we show in Fig.~\ref{fig:sector_stats} the sector statistics, which measures the fraction of time the impurity spends in a given atomic state. One can see that for the lowest filling, the impurity essentially is either completely empty ($|0000\rangle$) or singly occupied ($|1000\rangle$, $|0100\rangle$, $|0010\rangle$ or $|0001\rangle$). Doubly or higher occupied states are unlikely, so correlation effects are weak. Indeed, at these parameters the system has previously been identified as a weakly correlated Fermi liquid \cite{Hafermann12}. 
For a higher filling, $\langle n\rangle=1.29$, the system is on the boundary between the Fermi liquid and the frozen moment phase. While the probability of the singly occupied states remains almost unchanged, the contribution of the empty state becomes small, while the doubly occupied high-spin states ($|1010\rangle$ and $|0101\rangle$) become significantly populated. This indicates the formation of the spin $S=1$ moment. The low-spin doubly occupied states ($|1100\rangle$ and $|0011\rangle$) are suppressed due to the large intra-orbital Coulomb repulsion $U$, while $|0110\rangle$ and $|1001\rangle$ are suppressed due to $U'$. The latter have higher weight since $U'<U$.
For even higher filling ($\langle n\rangle$=1.53), deep in the frozen moment phase, the high-spin states gain further importance at the expense of the singly occupied states.
Additional results for this model can be found in Ref. \cite{Hafermann12}.

\section{Installation and Usage}
\subsection{Installation}
The program is provided as part of the ALPS \cite{ALPS20} package and can be downloaded from \href{http://alps.comp-phys.org}{http://alps.comp-phys.org}. Release milestones and binary nightly builds as well as the source code are available as a package and via anonymous svn access. The code binary \verb#hybridization# can be found in the ALPS binary directory \verb#alps_root/# \verb#bin#. Detailed documentation with an overview of all parameters and example scripts is located in \verb#alps_root/#\verb#doc# as \verb#hybdoc.pdf# .

\subsection{Usage}
The code is delivered together with tutorial examples which illustrate the usage of the program and allow one to reproduce the data presented here.
For a quick inspection and test of the program, do the following. After installation of the library
\begin{itemize}
\item From the directory \verb#alps_root/tutorials# enter directory \verb#hybridization-01-python#.
\item Inspect the file \verb#tutorial1.py#, which describes the parameters.
\item Run \verb#alps_root/bin/alpspython tutorial1.py#. This run will take about 30 seconds.
\item Plot \verb#Gt.dat# and compare it to \verb#Gt.dat# in the directory \verb#exact_diagonalization#, which contains exact results.
\item Plot the file \verb#orders.dat#, which contains the expansion orders.
\end{itemize}
For a more extensive look at the program, do the following.
\begin{itemize}
\item Enter directory \verb#hybridization-02-kondo#.
\item Inspect the file \verb#tutorial2.py#, which describes the parameters.
\item Run \verb#alps_root/bin/alpspython tutorial2.py#. The script will start several consecutive runs which each take a few seconds. After completion, a plot appears. Compare this plot to Fig.~\ref{fig:chiT}.
\end{itemize}
Two more tutorials generate the data in Figs.~\ref{fig:retint}-\ref{fig:sector_stats}. The extensive documentation in the subdirectory \verb#doc# provides more information and additional examples.

\section{Conclusions}

In conclusion, we provide here a state-of-the-art implementation of the hybridization expansion CT-QMC method which takes into account recent advances in methodology. Our code package also includes a set of examples that provide a pedagogical introduction, as well as Python scripts to reproduce the examples in this paper, and allow the user to test the implementation.
The code has been written specifically with users from the LDA+DMFT and nano-science community in mind, with the aim of providing both a starting point for students and a high-performance implementation that can be used in a production environment.

\section{Acknowledgments}
P.~W. acknowledges support by DFG FOR 1346 and SNF Grant 200021\_140648.
We gratefully acknowledge support by the wider ALPS \cite{ALPS20,ALPS05,ALPS07} community.

\bibliographystyle{apsrev4-1}
\bibliography{refs_shortened.bib}

%merlin.mbs apsrev4-1.bst 2010-07-25 4.21a (PWD, AO, DPC) hacked
%Control: key (0)
%Control: author (72) initials jnrlst
%Control: editor formatted (1) identically to author
%Control: production of article title (-1) disabled
%Control: page (0) single
%Control: year (1) truncated
%Control: production of eprint (0) enabled
\begin{thebibliography}{41}%
\makeatletter
\providecommand \@ifxundefined [1]{%
 \@ifx{#1\undefined}
}%
\providecommand \@ifnum [1]{%
 \ifnum #1\expandafter \@firstoftwo
 \else \expandafter \@secondoftwo
 \fi
}%
\providecommand \@ifx [1]{%
 \ifx #1\expandafter \@firstoftwo
 \else \expandafter \@secondoftwo
 \fi
}%
\providecommand \natexlab [1]{#1}%
\providecommand \enquote  [1]{``#1''}%
\providecommand \bibnamefont  [1]{#1}%
\providecommand \bibfnamefont [1]{#1}%
\providecommand \citenamefont [1]{#1}%
\providecommand \href@noop [0]{\@secondoftwo}%
\providecommand \href [0]{\begingroup \@sanitize@url \@href}%
\providecommand \@href[1]{\@@startlink{#1}\@@href}%
\providecommand \@@href[1]{\endgroup#1\@@endlink}%
\providecommand \@sanitize@url [0]{\catcode `\\12\catcode `\$12\catcode
  `\&12\catcode `\#12\catcode `\^12\catcode `\_12\catcode `\%12\relax}%
\providecommand \@@startlink[1]{}%
\providecommand \@@endlink[0]{}%
\providecommand \url  [0]{\begingroup\@sanitize@url \@url }%
\providecommand \@url [1]{\endgroup\@href {#1}{\urlprefix }}%
\providecommand \urlprefix  [0]{URL }%
\providecommand \Eprint [0]{\href }%
\providecommand \doibase [0]{http://dx.doi.org/}%
\providecommand \selectlanguage [0]{\@gobble}%
\providecommand \bibinfo  [0]{\@secondoftwo}%
\providecommand \bibfield  [0]{\@secondoftwo}%
\providecommand \translation [1]{[#1]}%
\providecommand \BibitemOpen [0]{}%
\providecommand \bibitemStop [0]{}%
\providecommand \bibitemNoStop [0]{.\EOS\space}%
\providecommand \EOS [0]{\spacefactor3000\relax}%
\providecommand \BibitemShut  [1]{\csname bibitem#1\endcsname}%
\let\auto@bib@innerbib\@empty
%</preamble>
\bibitem [{\citenamefont {Bauer}\ \emph {et~al.}(2011)\citenamefont {Bauer},
  \citenamefont {Carr}, \citenamefont {Evertz}, \citenamefont {Feiguin},
  \citenamefont {Freire}, \citenamefont {Fuchs}, \citenamefont {Gamper},
  \citenamefont {Gukelberger}, \citenamefont {Gull}, \citenamefont {Guertler},
  \citenamefont {Hehn}, \citenamefont {Igarashi}, \citenamefont {Isakov},
  \citenamefont {Koop}, \citenamefont {Ma}, \citenamefont {Mates},
  \citenamefont {Matsuo}, \citenamefont {Parcollet}, \citenamefont {Pawlowski},
  \citenamefont {Picon}, \citenamefont {Pollet}, \citenamefont {Santos},
  \citenamefont {Scarola}, \citenamefont {Schollw\"{o}ck}, \citenamefont
  {Silva}, \citenamefont {Surer}, \citenamefont {Todo}, \citenamefont {Trebst},
  \citenamefont {Troyer}, \citenamefont {Wall}, \citenamefont {Werner},\ and\
  \citenamefont {Wessel}}]{ALPS20}%
  \BibitemOpen
  \bibfield  {author} {\bibinfo {author} {\bibfnamefont {B.}~\bibnamefont
  {Bauer}}, \bibinfo {author} {\bibfnamefont {L.~D.}\ \bibnamefont {Carr}},
  \bibinfo {author} {\bibfnamefont {H.~G.}\ \bibnamefont {Evertz}}, \bibinfo
  {author} {\bibfnamefont {A.}~\bibnamefont {Feiguin}}, \bibinfo {author}
  {\bibfnamefont {J.}~\bibnamefont {Freire}}, \bibinfo {author} {\bibfnamefont
  {S.}~\bibnamefont {Fuchs}}, \bibinfo {author} {\bibfnamefont
  {L.}~\bibnamefont {Gamper}}, \bibinfo {author} {\bibfnamefont
  {J.}~\bibnamefont {Gukelberger}}, \bibinfo {author} {\bibfnamefont
  {E.}~\bibnamefont {Gull}}, \bibinfo {author} {\bibfnamefont {S.}~\bibnamefont
  {Guertler}}, \bibinfo {author} {\bibfnamefont {A.}~\bibnamefont {Hehn}},
  \bibinfo {author} {\bibfnamefont {R.}~\bibnamefont {Igarashi}}, \bibinfo
  {author} {\bibfnamefont {S.~V.}\ \bibnamefont {Isakov}}, \bibinfo {author}
  {\bibfnamefont {D.}~\bibnamefont {Koop}}, \bibinfo {author} {\bibfnamefont
  {P.~N.}\ \bibnamefont {Ma}}, \bibinfo {author} {\bibfnamefont
  {P.}~\bibnamefont {Mates}}, \bibinfo {author} {\bibfnamefont
  {H.}~\bibnamefont {Matsuo}}, \bibinfo {author} {\bibfnamefont
  {O.}~\bibnamefont {Parcollet}}, \bibinfo {author} {\bibfnamefont
  {G.}~\bibnamefont {Pawlowski}}, \bibinfo {author} {\bibfnamefont {J.~D.}\
  \bibnamefont {Picon}}, \bibinfo {author} {\bibfnamefont {L.}~\bibnamefont
  {Pollet}}, \bibinfo {author} {\bibfnamefont {E.}~\bibnamefont {Santos}},
  \bibinfo {author} {\bibfnamefont {V.~W.}\ \bibnamefont {Scarola}}, \bibinfo
  {author} {\bibfnamefont {U.}~\bibnamefont {Schollw\"{o}ck}}, \bibinfo
  {author} {\bibfnamefont {C.}~\bibnamefont {Silva}}, \bibinfo {author}
  {\bibfnamefont {B.}~\bibnamefont {Surer}}, \bibinfo {author} {\bibfnamefont
  {S.}~\bibnamefont {Todo}}, \bibinfo {author} {\bibfnamefont {S.}~\bibnamefont
  {Trebst}}, \bibinfo {author} {\bibfnamefont {M.}~\bibnamefont {Troyer}},
  \bibinfo {author} {\bibfnamefont {M.~L.}\ \bibnamefont {Wall}}, \bibinfo
  {author} {\bibfnamefont {P.}~\bibnamefont {Werner}}, \ and\ \bibinfo {author}
  {\bibfnamefont {S.}~\bibnamefont {Wessel}},\ }\href@noop {} {\bibfield
  {journal} {\bibinfo  {journal} {Journal of Statistical Mechanics: Theory and
  Experiment}\ }\textbf {\bibinfo {volume} {2011}},\ \bibinfo {pages} {P05001}
  (\bibinfo {year} {2011})}\BibitemShut {NoStop}%
\bibitem [{\citenamefont {Alet}\ \emph {et~al.}(2005)\citenamefont {Alet},
  \citenamefont {Dayal}, \citenamefont {Grzesik}, \citenamefont {Honecker},
  \citenamefont {K\"{o}rner}, \citenamefont {L\"{a}uchli}, \citenamefont
  {Manmana}, \citenamefont {McCulloch}, \citenamefont {Michel}, \citenamefont
  {Noack}, \citenamefont {Schmid}, \citenamefont {Schollw\"{o}ck},
  \citenamefont {St\"{o}ckli}, \citenamefont {Todo}, \citenamefont {Trebst},
  \citenamefont {Troyer}, \citenamefont {Werner},\ and\ \citenamefont
  {collaboration}}]{ALPS05}%
  \BibitemOpen
  \bibfield  {author} {\bibinfo {author} {\bibfnamefont {F.}~\bibnamefont
  {Alet}}, \bibinfo {author} {\bibfnamefont {P.}~\bibnamefont {Dayal}},
  \bibinfo {author} {\bibfnamefont {A.}~\bibnamefont {Grzesik}}, \bibinfo
  {author} {\bibfnamefont {A.}~\bibnamefont {Honecker}}, \bibinfo {author}
  {\bibfnamefont {M.}~\bibnamefont {K\"{o}rner}}, \bibinfo {author}
  {\bibfnamefont {A.}~\bibnamefont {L\"{a}uchli}}, \bibinfo {author}
  {\bibfnamefont {S.~R.}\ \bibnamefont {Manmana}}, \bibinfo {author}
  {\bibfnamefont {I.~P.}\ \bibnamefont {McCulloch}}, \bibinfo {author}
  {\bibfnamefont {F.}~\bibnamefont {Michel}}, \bibinfo {author} {\bibfnamefont
  {R.~M.}\ \bibnamefont {Noack}}, \bibinfo {author} {\bibfnamefont
  {G.}~\bibnamefont {Schmid}}, \bibinfo {author} {\bibfnamefont
  {U.}~\bibnamefont {Schollw\"{o}ck}}, \bibinfo {author} {\bibfnamefont
  {F.}~\bibnamefont {St\"{o}ckli}}, \bibinfo {author} {\bibfnamefont
  {S.}~\bibnamefont {Todo}}, \bibinfo {author} {\bibfnamefont {S.}~\bibnamefont
  {Trebst}}, \bibinfo {author} {\bibfnamefont {M.}~\bibnamefont {Troyer}},
  \bibinfo {author} {\bibfnamefont {P.}~\bibnamefont {Werner}}, \ and\ \bibinfo
  {author} {\bibfnamefont {S.~W.}\ \bibnamefont {collaboration}},\ }\href
  {\doibase 10.1143/JPSJS.74S.30} {\bibfield  {journal} {\bibinfo  {journal}
  {J. Phys. Soc. Jpn.}\ }\textbf {\bibinfo {volume} {74S}},\ \bibinfo {pages}
  {30} (\bibinfo {year} {2005})}\BibitemShut {NoStop}%
\bibitem [{\citenamefont {Albuquerque}\ \emph {et~al.}(2007)\citenamefont
  {Albuquerque}, \citenamefont {Alet}, \citenamefont {Corboz}, \citenamefont
  {Dayal}, \citenamefont {Feiguin}, \citenamefont {Fuchs}, \citenamefont
  {Gamper}, \citenamefont {Gull}, \citenamefont {G\"{u}rtler}, \citenamefont
  {Honecker}, \citenamefont {Igarashi}, \citenamefont {K\"{o}rner},
  \citenamefont {Kozhevnikov}, \citenamefont {L\"{a}uchli}, \citenamefont
  {Manmana}, \citenamefont {Matsumoto}, \citenamefont {McCulloch},
  \citenamefont {Michel}, \citenamefont {Noack}, \citenamefont {Pawlowski},
  \citenamefont {Pollet}, \citenamefont {Pruschke}, \citenamefont
  {Schollw\"{o}ck}, \citenamefont {Todo}, \citenamefont {Trebst}, \citenamefont
  {Troyer}, \citenamefont {Werner},\ and\ \citenamefont {Wessel}}]{ALPS07}%
  \BibitemOpen
  \bibfield  {author} {\bibinfo {author} {\bibfnamefont {A.}~\bibnamefont
  {Albuquerque}}, \bibinfo {author} {\bibfnamefont {F.}~\bibnamefont {Alet}},
  \bibinfo {author} {\bibfnamefont {P.}~\bibnamefont {Corboz}}, \bibinfo
  {author} {\bibfnamefont {P.}~\bibnamefont {Dayal}}, \bibinfo {author}
  {\bibfnamefont {A.}~\bibnamefont {Feiguin}}, \bibinfo {author} {\bibfnamefont
  {S.}~\bibnamefont {Fuchs}}, \bibinfo {author} {\bibfnamefont
  {L.}~\bibnamefont {Gamper}}, \bibinfo {author} {\bibfnamefont
  {E.}~\bibnamefont {Gull}}, \bibinfo {author} {\bibfnamefont {S.}~\bibnamefont
  {G\"{u}rtler}}, \bibinfo {author} {\bibfnamefont {A.}~\bibnamefont
  {Honecker}}, \bibinfo {author} {\bibfnamefont {R.}~\bibnamefont {Igarashi}},
  \bibinfo {author} {\bibfnamefont {M.}~\bibnamefont {K\"{o}rner}}, \bibinfo
  {author} {\bibfnamefont {A.}~\bibnamefont {Kozhevnikov}}, \bibinfo {author}
  {\bibfnamefont {A.}~\bibnamefont {L\"{a}uchli}}, \bibinfo {author}
  {\bibfnamefont {S.}~\bibnamefont {Manmana}}, \bibinfo {author} {\bibfnamefont
  {M.}~\bibnamefont {Matsumoto}}, \bibinfo {author} {\bibfnamefont
  {I.}~\bibnamefont {McCulloch}}, \bibinfo {author} {\bibfnamefont
  {F.}~\bibnamefont {Michel}}, \bibinfo {author} {\bibfnamefont
  {R.}~\bibnamefont {Noack}}, \bibinfo {author} {\bibfnamefont
  {G.}~\bibnamefont {Pawlowski}}, \bibinfo {author} {\bibfnamefont
  {L.}~\bibnamefont {Pollet}}, \bibinfo {author} {\bibfnamefont
  {T.}~\bibnamefont {Pruschke}}, \bibinfo {author} {\bibfnamefont
  {U.}~\bibnamefont {Schollw\"{o}ck}}, \bibinfo {author} {\bibfnamefont
  {S.}~\bibnamefont {Todo}}, \bibinfo {author} {\bibfnamefont {S.}~\bibnamefont
  {Trebst}}, \bibinfo {author} {\bibfnamefont {M.}~\bibnamefont {Troyer}},
  \bibinfo {author} {\bibfnamefont {P.}~\bibnamefont {Werner}}, \ and\ \bibinfo
  {author} {\bibfnamefont {S.}~\bibnamefont {Wessel}},\ }\href {\doibase
  10.1016/j.jmmm.2006.10.304} {\bibfield  {journal} {\bibinfo  {journal} {J.
  Magn. Magn. Mater.}\ }\textbf {\bibinfo {volume} {310}},\ \bibinfo {pages}
  {1187 } (\bibinfo {year} {2007})},\ \bibinfo {note} {proceedings of the 17th
  International Conference on Magnetism -- The International Conference on
  Magnetism}\BibitemShut {NoStop}%
\bibitem [{\citenamefont {Lawson}\ \emph {et~al.}(1979)\citenamefont {Lawson},
  \citenamefont {Hanson}, \citenamefont {Kincaid},\ and\ \citenamefont
  {Krogh}}]{BLAS79}%
  \BibitemOpen
  \bibfield  {author} {\bibinfo {author} {\bibfnamefont {C.~L.}\ \bibnamefont
  {Lawson}}, \bibinfo {author} {\bibfnamefont {R.~J.}\ \bibnamefont {Hanson}},
  \bibinfo {author} {\bibfnamefont {D.~R.}\ \bibnamefont {Kincaid}}, \ and\
  \bibinfo {author} {\bibfnamefont {F.~T.}\ \bibnamefont {Krogh}},\ }\href@noop
  {} {\bibfield  {journal} {\bibinfo  {journal} {{ACM} Transactions on
  Mathematical Software}\ }\textbf {\bibinfo {volume} {5}},\ \bibinfo {pages}
  {324} (\bibinfo {year} {1979})}\BibitemShut {NoStop}%
\bibitem [{\citenamefont {Blackford}\ \emph {et~al.}(2002)\citenamefont
  {Blackford}, \citenamefont {Demmel}, \citenamefont {Duff}, \citenamefont
  {Henry}, \citenamefont {Heroux}, \citenamefont {Kaufman}, \citenamefont
  {Lumsdaine}, \citenamefont {Petitet},\ and\ \citenamefont {Whaley}}]{BLAS02}%
  \BibitemOpen
  \bibfield  {author} {\bibinfo {author} {\bibfnamefont {L.~S.}\ \bibnamefont
  {Blackford}}, \bibinfo {author} {\bibfnamefont {J.}~\bibnamefont {Demmel}},
  \bibinfo {author} {\bibfnamefont {I.}~\bibnamefont {Duff}}, \bibinfo {author}
  {\bibfnamefont {G.}~\bibnamefont {Henry}}, \bibinfo {author} {\bibfnamefont
  {M.}~\bibnamefont {Heroux}}, \bibinfo {author} {\bibfnamefont
  {L.}~\bibnamefont {Kaufman}}, \bibinfo {author} {\bibfnamefont
  {A.}~\bibnamefont {Lumsdaine}}, \bibinfo {author} {\bibfnamefont
  {A.}~\bibnamefont {Petitet}}, \ and\ \bibinfo {author} {\bibfnamefont
  {R.~C.}\ \bibnamefont {Whaley}},\ }\href {\doibase 10.1145/567806.567807}
  {\bibfield  {journal} {\bibinfo  {journal} {ACM Trans. Math. Softw.}\
  }\textbf {\bibinfo {volume} {28}},\ \bibinfo {pages} {135} (\bibinfo {year}
  {2002})}\BibitemShut {NoStop}%
\bibitem [{\citenamefont {Anderson}\ \emph {et~al.}(1999)\citenamefont
  {Anderson}, \citenamefont {Bai}, \citenamefont {Bischof}, \citenamefont
  {Blackford}, \citenamefont {Demmel}, \citenamefont {Dongarra}, \citenamefont
  {Du~Croz}, \citenamefont {Greenbaum}, \citenamefont {Hammarling},
  \citenamefont {McKenney},\ and\ \citenamefont {Sorensen}}]{LAPACK99}%
  \BibitemOpen
  \bibfield  {author} {\bibinfo {author} {\bibfnamefont {E.}~\bibnamefont
  {Anderson}}, \bibinfo {author} {\bibfnamefont {Z.}~\bibnamefont {Bai}},
  \bibinfo {author} {\bibfnamefont {C.}~\bibnamefont {Bischof}}, \bibinfo
  {author} {\bibfnamefont {S.}~\bibnamefont {Blackford}}, \bibinfo {author}
  {\bibfnamefont {J.}~\bibnamefont {Demmel}}, \bibinfo {author} {\bibfnamefont
  {J.}~\bibnamefont {Dongarra}}, \bibinfo {author} {\bibfnamefont
  {J.}~\bibnamefont {Du~Croz}}, \bibinfo {author} {\bibfnamefont
  {A.}~\bibnamefont {Greenbaum}}, \bibinfo {author} {\bibfnamefont
  {S.}~\bibnamefont {Hammarling}}, \bibinfo {author} {\bibfnamefont
  {A.}~\bibnamefont {McKenney}}, \ and\ \bibinfo {author} {\bibfnamefont
  {D.}~\bibnamefont {Sorensen}},\ }\href@noop {} {\emph {\bibinfo {title}
  {{LAPACK} Users' Guide}}},\ \bibinfo {edition} {3rd}\ ed.\ (\bibinfo
  {publisher} {Society for Industrial and Applied Mathematics},\ \bibinfo
  {address} {Philadelphia, PA},\ \bibinfo {year} {1999})\BibitemShut {NoStop}%
\bibitem [{\citenamefont {{The HDF Group}}(2010)}]{hdf5}%
  \BibitemOpen
  \bibfield  {author} {\bibinfo {author} {\bibnamefont {{The HDF Group}}},\
  }\href@noop {} {\enquote {\bibinfo {title} {Hierarchical data format version
  5},}\ }\bibinfo {howpublished} {http://www.hdfgroup.org/HDF5} (\bibinfo
  {year} {2000-2010})\BibitemShut {NoStop}%
\bibitem [{\citenamefont {Anderson}(1961)}]{Anderson61}%
  \BibitemOpen
  \bibfield  {author} {\bibinfo {author} {\bibfnamefont {P.~W.}\ \bibnamefont
  {Anderson}},\ }\href {\doibase 10.1103/PhysRev.124.41} {\bibfield  {journal}
  {\bibinfo  {journal} {Phys. Rev.}\ }\textbf {\bibinfo {volume} {124}},\
  \bibinfo {pages} {41} (\bibinfo {year} {1961})}\BibitemShut {NoStop}%
\bibitem [{\citenamefont {Hanson}\ \emph {et~al.}(2007)\citenamefont {Hanson},
  \citenamefont {Kouwenhoven}, \citenamefont {Petta}, \citenamefont {Tarucha},\
  and\ \citenamefont {Vandersypen}}]{Hanson07}%
  \BibitemOpen
  \bibfield  {author} {\bibinfo {author} {\bibfnamefont {R.}~\bibnamefont
  {Hanson}}, \bibinfo {author} {\bibfnamefont {L.~P.}\ \bibnamefont
  {Kouwenhoven}}, \bibinfo {author} {\bibfnamefont {J.~R.}\ \bibnamefont
  {Petta}}, \bibinfo {author} {\bibfnamefont {S.}~\bibnamefont {Tarucha}}, \
  and\ \bibinfo {author} {\bibfnamefont {L.~M.~K.}\ \bibnamefont
  {Vandersypen}},\ }\href {\doibase 10.1103/RevModPhys.79.1217} {\bibfield
  {journal} {\bibinfo  {journal} {Rev. Mod. Phys.}\ }\textbf {\bibinfo {volume}
  {79}},\ \bibinfo {eid} {1217} (\bibinfo {year} {2007})}\BibitemShut {NoStop}%
\bibitem [{\citenamefont {Brako}\ and\ \citenamefont {Newns}(1981)}]{Brako81}%
  \BibitemOpen
  \bibfield  {author} {\bibinfo {author} {\bibfnamefont {R.}~\bibnamefont
  {Brako}}\ and\ \bibinfo {author} {\bibfnamefont {D.~M.}\ \bibnamefont
  {Newns}},\ }\href@noop {} {\bibfield  {journal} {\bibinfo  {journal} {Journal
  of Physics C: Solid State Physics}\ }\textbf {\bibinfo {volume} {14}},\
  \bibinfo {pages} {3065} (\bibinfo {year} {1981})}\BibitemShut {NoStop}%
\bibitem [{\citenamefont {Metzner}\ and\ \citenamefont
  {Vollhardt}(1989)}]{Metzner89}%
  \BibitemOpen
  \bibfield  {author} {\bibinfo {author} {\bibfnamefont {W.}~\bibnamefont
  {Metzner}}\ and\ \bibinfo {author} {\bibfnamefont {D.}~\bibnamefont
  {Vollhardt}},\ }\href {\doibase 10.1103/PhysRevLett.62.324} {\bibfield
  {journal} {\bibinfo  {journal} {Phys. Rev. Lett.}\ }\textbf {\bibinfo
  {volume} {62}},\ \bibinfo {pages} {324} (\bibinfo {year} {1989})}\BibitemShut
  {NoStop}%
\bibitem [{\citenamefont {Georges}\ and\ \citenamefont
  {Krauth}(1992)}]{Georges92}%
  \BibitemOpen
  \bibfield  {author} {\bibinfo {author} {\bibfnamefont {A.}~\bibnamefont
  {Georges}}\ and\ \bibinfo {author} {\bibfnamefont {W.}~\bibnamefont
  {Krauth}},\ }\href {\doibase 10.1103/PhysRevLett.69.1240} {\bibfield
  {journal} {\bibinfo  {journal} {Phys. Rev. Lett.}\ }\textbf {\bibinfo
  {volume} {69}},\ \bibinfo {pages} {1240} (\bibinfo {year}
  {1992})}\BibitemShut {NoStop}%
\bibitem [{\citenamefont {Georges}\ \emph {et~al.}(1996)\citenamefont
  {Georges}, \citenamefont {Kotliar}, \citenamefont {Krauth},\ and\
  \citenamefont {Rozenberg}}]{Georges96}%
  \BibitemOpen
  \bibfield  {author} {\bibinfo {author} {\bibfnamefont {A.}~\bibnamefont
  {Georges}}, \bibinfo {author} {\bibfnamefont {G.}~\bibnamefont {Kotliar}},
  \bibinfo {author} {\bibfnamefont {W.}~\bibnamefont {Krauth}}, \ and\ \bibinfo
  {author} {\bibfnamefont {M.~J.}\ \bibnamefont {Rozenberg}},\ }\href {\doibase
  10.1103/RevModPhys.68.13} {\bibfield  {journal} {\bibinfo  {journal} {Rev.
  Mod. Phys.}\ }\textbf {\bibinfo {volume} {68}},\ \bibinfo {pages} {13}
  (\bibinfo {year} {1996})}\BibitemShut {NoStop}%
\bibitem [{\citenamefont {Georges}(2004)}]{Georges04}%
  \BibitemOpen
  \bibfield  {author} {\bibinfo {author} {\bibfnamefont {A.}~\bibnamefont
  {Georges}},\ }\href {\doibase 10.1063/1.1800733} {\bibfield  {journal}
  {\bibinfo  {journal} {LECTURES ON THE PHYSICS OF HIGHLY CORRELATED ELECTRON
  SYSTEMS VIII: Eighth Training Course in the Physics of Correlated Electron
  Systems and High-Tc Superconductors}\ }\textbf {\bibinfo {volume} {715}},\
  \bibinfo {pages} {3} (\bibinfo {year} {2004})}\BibitemShut {NoStop}%
\bibitem [{\citenamefont {Held}\ \emph {et~al.}(2006)\citenamefont {Held},
  \citenamefont {Nekrasov}, \citenamefont {Keller}, \citenamefont {Eyert},
  \citenamefont {Bluemer}, \citenamefont {McMahan}, \citenamefont {Scalettar},
  \citenamefont {Pruschke}, \citenamefont {Anisimov},\ and\ \citenamefont
  {Vollhardt}}]{Held06}%
  \BibitemOpen
  \bibfield  {author} {\bibinfo {author} {\bibfnamefont {K.}~\bibnamefont
  {Held}}, \bibinfo {author} {\bibfnamefont {I.~A.}\ \bibnamefont {Nekrasov}},
  \bibinfo {author} {\bibfnamefont {G.}~\bibnamefont {Keller}}, \bibinfo
  {author} {\bibfnamefont {V.}~\bibnamefont {Eyert}}, \bibinfo {author}
  {\bibfnamefont {N.}~\bibnamefont {Bluemer}}, \bibinfo {author} {\bibfnamefont
  {A.~K.}\ \bibnamefont {McMahan}}, \bibinfo {author} {\bibfnamefont {R.~T.}\
  \bibnamefont {Scalettar}}, \bibinfo {author} {\bibfnamefont {T.}~\bibnamefont
  {Pruschke}}, \bibinfo {author} {\bibfnamefont {V.~I.}\ \bibnamefont
  {Anisimov}}, \ and\ \bibinfo {author} {\bibfnamefont {D.}~\bibnamefont
  {Vollhardt}},\ }\href {\doibase 10.1002/pssb.200642053} {\bibfield  {journal}
  {\bibinfo  {journal} {Phys. Status Solidi}\ }\textbf {\bibinfo {volume}
  {243}},\ \bibinfo {pages} {2599} (\bibinfo {year} {2006})}\BibitemShut
  {NoStop}%
\bibitem [{\citenamefont {Held}(2007)}]{Held07}%
  \BibitemOpen
  \bibfield  {author} {\bibinfo {author} {\bibfnamefont {K.}~\bibnamefont
  {Held}},\ }\href {\doibase 10.1080/00018730701619647} {\bibfield  {journal}
  {\bibinfo  {journal} {Advances in Physics}\ }\textbf {\bibinfo {volume}
  {56}},\ \bibinfo {pages} {829} (\bibinfo {year} {2007})}\BibitemShut
  {NoStop}%
\bibitem [{\citenamefont {Kotliar}\ \emph {et~al.}(2006)\citenamefont
  {Kotliar}, \citenamefont {Savrasov}, \citenamefont {Haule} \emph
  {et~al.}}]{Kotliar06}%
  \BibitemOpen
  \bibfield  {author} {\bibinfo {author} {\bibfnamefont {G.}~\bibnamefont
  {Kotliar}}, \bibinfo {author} {\bibfnamefont {S.~Y.}\ \bibnamefont
  {Savrasov}}, \bibinfo {author} {\bibfnamefont {K.}~\bibnamefont {Haule}},
  \emph {et~al.},\ }\href {\doibase 10.1103/RevModPhys.78.865} {\bibfield
  {journal} {\bibinfo  {journal} {Rev. Mod. Phys.}\ }\textbf {\bibinfo {volume}
  {78}},\ \bibinfo {eid} {865} (\bibinfo {year} {2006})}\BibitemShut {NoStop}%
\bibitem [{\citenamefont {Maier}\ \emph {et~al.}(2005)\citenamefont {Maier},
  \citenamefont {Jarrell}, \citenamefont {Pruschke},\ and\ \citenamefont
  {Hettler}}]{Maier05}%
  \BibitemOpen
  \bibfield  {author} {\bibinfo {author} {\bibfnamefont {T.}~\bibnamefont
  {Maier}}, \bibinfo {author} {\bibfnamefont {M.}~\bibnamefont {Jarrell}},
  \bibinfo {author} {\bibfnamefont {T.}~\bibnamefont {Pruschke}}, \ and\
  \bibinfo {author} {\bibfnamefont {M.~H.}\ \bibnamefont {Hettler}},\ }\href
  {\doibase 10.1103/RevModPhys.77.1027} {\bibfield  {journal} {\bibinfo
  {journal} {Rev. Mod. Phys.}\ }\textbf {\bibinfo {volume} {77}},\ \bibinfo
  {eid} {1027} (\bibinfo {year} {2005})}\BibitemShut {NoStop}%
\bibitem [{\citenamefont {Rubtsov}\ and\ \citenamefont
  {Lichtenstein}(2004)}]{Rubtsov04}%
  \BibitemOpen
  \bibfield  {author} {\bibinfo {author} {\bibfnamefont {A.~N.}\ \bibnamefont
  {Rubtsov}}\ and\ \bibinfo {author} {\bibfnamefont {A.~I.}\ \bibnamefont
  {Lichtenstein}},\ }\href {\doibase 10.1134/1.1800216} {\bibfield  {journal}
  {\bibinfo  {journal} {JETP Letters}\ }\textbf {\bibinfo {volume} {80}},\
  \bibinfo {pages} {61} (\bibinfo {year} {2004})}\BibitemShut {NoStop}%
\bibitem [{\citenamefont {Rubtsov}\ \emph {et~al.}(2005)\citenamefont
  {Rubtsov}, \citenamefont {Savkin},\ and\ \citenamefont
  {Lichtenstein}}]{Rubtsov05}%
  \BibitemOpen
  \bibfield  {author} {\bibinfo {author} {\bibfnamefont {A.~N.}\ \bibnamefont
  {Rubtsov}}, \bibinfo {author} {\bibfnamefont {V.~V.}\ \bibnamefont {Savkin}},
  \ and\ \bibinfo {author} {\bibfnamefont {A.~I.}\ \bibnamefont
  {Lichtenstein}},\ }\href {\doibase 10.1103/PhysRevB.72.035122} {\bibfield
  {journal} {\bibinfo  {journal} {Phys. Rev. B}\ }\textbf {\bibinfo {volume}
  {72}},\ \bibinfo {eid} {035122} (\bibinfo {year} {2005})}\BibitemShut
  {NoStop}%
\bibitem [{\citenamefont {Gull}\ \emph
  {et~al.}(2011{\natexlab{a}})\citenamefont {Gull}, \citenamefont {Millis},
  \citenamefont {Lichtenstein}, \citenamefont {Rubtsov}, \citenamefont
  {Troyer},\ and\ \citenamefont {Werner}}]{Gull11_review}%
  \BibitemOpen
  \bibfield  {author} {\bibinfo {author} {\bibfnamefont {E.}~\bibnamefont
  {Gull}}, \bibinfo {author} {\bibfnamefont {A.~J.}\ \bibnamefont {Millis}},
  \bibinfo {author} {\bibfnamefont {A.~I.}\ \bibnamefont {Lichtenstein}},
  \bibinfo {author} {\bibfnamefont {A.~N.}\ \bibnamefont {Rubtsov}}, \bibinfo
  {author} {\bibfnamefont {M.}~\bibnamefont {Troyer}}, \ and\ \bibinfo {author}
  {\bibfnamefont {P.}~\bibnamefont {Werner}},\ }\href {\doibase
  10.1103/RevModPhys.83.349} {\bibfield  {journal} {\bibinfo  {journal} {Rev.
  Mod. Phys.}\ }\textbf {\bibinfo {volume} {83}},\ \bibinfo {pages} {349}
  (\bibinfo {year} {2011}{\natexlab{a}})}\BibitemShut {NoStop}%
\bibitem [{\citenamefont {Werner}\ \emph {et~al.}(2006)\citenamefont {Werner},
  \citenamefont {Comanac}, \citenamefont {de' Medici} \emph
  {et~al.}}]{Werner06}%
  \BibitemOpen
  \bibfield  {author} {\bibinfo {author} {\bibfnamefont {P.}~\bibnamefont
  {Werner}}, \bibinfo {author} {\bibfnamefont {A.}~\bibnamefont {Comanac}},
  \bibinfo {author} {\bibfnamefont {L.}~\bibnamefont {de' Medici}},  \emph
  {et~al.},\ }\href {\doibase 10.1103/PhysRevLett.97.076405} {\bibfield
  {journal} {\bibinfo  {journal} {Phys. Rev. Lett.}\ }\textbf {\bibinfo
  {volume} {97}},\ \bibinfo {eid} {076405} (\bibinfo {year}
  {2006})}\BibitemShut {NoStop}%
\bibitem [{\citenamefont {Hafermann}\ \emph {et~al.}(2012)\citenamefont
  {Hafermann}, \citenamefont {Patton},\ and\ \citenamefont
  {Werner}}]{Hafermann12}%
  \BibitemOpen
  \bibfield  {author} {\bibinfo {author} {\bibfnamefont {H.}~\bibnamefont
  {Hafermann}}, \bibinfo {author} {\bibfnamefont {K.~R.}\ \bibnamefont
  {Patton}}, \ and\ \bibinfo {author} {\bibfnamefont {P.}~\bibnamefont
  {Werner}},\ }\href {\doibase 10.1103/PhysRevB.85.205106} {\bibfield
  {journal} {\bibinfo  {journal} {Phys. Rev. B}\ }\textbf {\bibinfo {volume}
  {85}},\ \bibinfo {pages} {205106} (\bibinfo {year} {2012})}\BibitemShut
  {NoStop}%
\bibitem [{\citenamefont {Boehnke}\ \emph {et~al.}(2011)\citenamefont
  {Boehnke}, \citenamefont {Hafermann}, \citenamefont {Ferrero}, \citenamefont
  {Lechermann},\ and\ \citenamefont {Parcollet}}]{Boehnke11}%
  \BibitemOpen
  \bibfield  {author} {\bibinfo {author} {\bibfnamefont {L.}~\bibnamefont
  {Boehnke}}, \bibinfo {author} {\bibfnamefont {H.}~\bibnamefont {Hafermann}},
  \bibinfo {author} {\bibfnamefont {M.}~\bibnamefont {Ferrero}}, \bibinfo
  {author} {\bibfnamefont {F.}~\bibnamefont {Lechermann}}, \ and\ \bibinfo
  {author} {\bibfnamefont {O.}~\bibnamefont {Parcollet}},\ }\href {\doibase
  10.1103/PhysRevB.84.075145} {\bibfield  {journal} {\bibinfo  {journal} {Phys.
  Rev. B}\ }\textbf {\bibinfo {volume} {84}},\ \bibinfo {pages} {075145}
  (\bibinfo {year} {2011})}\BibitemShut {NoStop}%
\bibitem [{\citenamefont {Werner}\ and\ \citenamefont
  {Millis}(2007)}]{Werner07holstein}%
  \BibitemOpen
  \bibfield  {author} {\bibinfo {author} {\bibfnamefont {P.}~\bibnamefont
  {Werner}}\ and\ \bibinfo {author} {\bibfnamefont {A.~J.}\ \bibnamefont
  {Millis}},\ }\href {\doibase 10.1103/PhysRevLett.99.146404} {\bibfield
  {journal} {\bibinfo  {journal} {Phys. Rev. Lett.}\ }\textbf {\bibinfo
  {volume} {99}},\ \bibinfo {eid} {146404} (\bibinfo {year}
  {2007})}\BibitemShut {NoStop}%
\bibitem [{\citenamefont {Werner}\ and\ \citenamefont
  {Millis}(2010)}]{Werner10_frequency}%
  \BibitemOpen
  \bibfield  {author} {\bibinfo {author} {\bibfnamefont {P.}~\bibnamefont
  {Werner}}\ and\ \bibinfo {author} {\bibfnamefont {A.~J.}\ \bibnamefont
  {Millis}},\ }\href {\doibase 10.1103/PhysRevLett.104.146401} {\bibfield
  {journal} {\bibinfo  {journal} {Phys. Rev. Lett.}\ }\textbf {\bibinfo
  {volume} {104}},\ \bibinfo {pages} {146401} (\bibinfo {year}
  {2010})}\BibitemShut {NoStop}%
\bibitem [{\citenamefont {Gull}\ \emph
  {et~al.}(2011{\natexlab{b}})\citenamefont {Gull}, \citenamefont {Staar},
  \citenamefont {Fuchs}, \citenamefont {Nukala}, \citenamefont {Summers},
  \citenamefont {Pruschke}, \citenamefont {Schulthess},\ and\ \citenamefont
  {Maier}}]{Gull10_submatrix}%
  \BibitemOpen
  \bibfield  {author} {\bibinfo {author} {\bibfnamefont {E.}~\bibnamefont
  {Gull}}, \bibinfo {author} {\bibfnamefont {P.}~\bibnamefont {Staar}},
  \bibinfo {author} {\bibfnamefont {S.}~\bibnamefont {Fuchs}}, \bibinfo
  {author} {\bibfnamefont {P.}~\bibnamefont {Nukala}}, \bibinfo {author}
  {\bibfnamefont {M.~S.}\ \bibnamefont {Summers}}, \bibinfo {author}
  {\bibfnamefont {T.}~\bibnamefont {Pruschke}}, \bibinfo {author}
  {\bibfnamefont {T.~C.}\ \bibnamefont {Schulthess}}, \ and\ \bibinfo {author}
  {\bibfnamefont {T.}~\bibnamefont {Maier}},\ }\href {\doibase
  10.1103/PhysRevB.83.075122} {\bibfield  {journal} {\bibinfo  {journal} {Phys.
  Rev. B}\ }\textbf {\bibinfo {volume} {83}},\ \bibinfo {pages} {075122}
  (\bibinfo {year} {2011}{\natexlab{b}})}\BibitemShut {NoStop}%
\bibitem [{\citenamefont {Werner}\ and\ \citenamefont
  {Millis}(2006)}]{Werner06Kondo}%
  \BibitemOpen
  \bibfield  {author} {\bibinfo {author} {\bibfnamefont {P.}~\bibnamefont
  {Werner}}\ and\ \bibinfo {author} {\bibfnamefont {A.~J.}\ \bibnamefont
  {Millis}},\ }\href {\doibase 10.1103/PhysRevB.74.155107} {\bibfield
  {journal} {\bibinfo  {journal} {Phys. Rev. B}\ }\textbf {\bibinfo {volume}
  {74}},\ \bibinfo {eid} {155107} (\bibinfo {year} {2006})}\BibitemShut
  {NoStop}%
\bibitem [{\citenamefont {Haule}(2007)}]{Haule07}%
  \BibitemOpen
  \bibfield  {author} {\bibinfo {author} {\bibfnamefont {K.}~\bibnamefont
  {Haule}},\ }\href {\doibase 10.1103/PhysRevB.75.155113} {\bibfield  {journal}
  {\bibinfo  {journal} {Phys. Rev. B}\ }\textbf {\bibinfo {volume} {75}},\
  \bibinfo {eid} {155113} (\bibinfo {year} {2007})}\BibitemShut {NoStop}%
\bibitem [{\citenamefont {Gull}\ \emph {et~al.}(2008)\citenamefont {Gull},
  \citenamefont {Werner}, \citenamefont {Parcollet},\ and\ \citenamefont
  {Troyer}}]{Gull08_ctaux}%
  \BibitemOpen
  \bibfield  {author} {\bibinfo {author} {\bibfnamefont {E.}~\bibnamefont
  {Gull}}, \bibinfo {author} {\bibfnamefont {P.}~\bibnamefont {Werner}},
  \bibinfo {author} {\bibfnamefont {O.}~\bibnamefont {Parcollet}}, \ and\
  \bibinfo {author} {\bibfnamefont {M.}~\bibnamefont {Troyer}},\ }\href@noop {}
  {\bibfield  {journal} {\bibinfo  {journal} {Europhys. Lett.}\ }\textbf
  {\bibinfo {volume} {82}},\ \bibinfo {pages} {57003 (6pp)} (\bibinfo {year}
  {2008})}\BibitemShut {NoStop}%
\bibitem [{\citenamefont {L\"auchli}\ and\ \citenamefont
  {Werner}(2009)}]{Laeuchli09}%
  \BibitemOpen
  \bibfield  {author} {\bibinfo {author} {\bibfnamefont {A.~M.}\ \bibnamefont
  {L\"auchli}}\ and\ \bibinfo {author} {\bibfnamefont {P.}~\bibnamefont
  {Werner}},\ }\href {\doibase 10.1103/PhysRevB.80.235117} {\bibfield
  {journal} {\bibinfo  {journal} {Phys. Rev. B}\ }\textbf {\bibinfo {volume}
  {80}},\ \bibinfo {pages} {235117} (\bibinfo {year} {2009})}\BibitemShut
  {NoStop}%
\bibitem [{\citenamefont {{Huang}}\ and\ \citenamefont
  {{Dai}}(2012)}]{Huang12b}%
  \BibitemOpen
  \bibfield  {author} {\bibinfo {author} {\bibfnamefont {L.}~\bibnamefont
  {{Huang}}}\ and\ \bibinfo {author} {\bibfnamefont {X.}~\bibnamefont
  {{Dai}}},\ }\href@noop {} {\bibfield  {journal} {\bibinfo  {journal} {ArXiv
  e-prints}\ } (\bibinfo {year} {2012})}\BibitemShut {NoStop}%
\bibitem [{\citenamefont {Parragh}\ \emph {et~al.}(2012)\citenamefont
  {Parragh}, \citenamefont {Toschi}, \citenamefont {Held},\ and\ \citenamefont
  {Sangiovanni}}]{Parragh12}%
  \BibitemOpen
  \bibfield  {author} {\bibinfo {author} {\bibfnamefont {N.}~\bibnamefont
  {Parragh}}, \bibinfo {author} {\bibfnamefont {A.}~\bibnamefont {Toschi}},
  \bibinfo {author} {\bibfnamefont {K.}~\bibnamefont {Held}}, \ and\ \bibinfo
  {author} {\bibfnamefont {G.}~\bibnamefont {Sangiovanni}},\ }\href {\doibase
  10.1103/PhysRevB.86.155158} {\bibfield  {journal} {\bibinfo  {journal} {Phys.
  Rev. B}\ }\textbf {\bibinfo {volume} {86}},\ \bibinfo {pages} {155158}
  (\bibinfo {year} {2012})}\BibitemShut {NoStop}%
\bibitem [{\citenamefont {Holstein}(1959)}]{Holstein59}%
  \BibitemOpen
  \bibfield  {author} {\bibinfo {author} {\bibfnamefont {T.}~\bibnamefont
  {Holstein}},\ }\href {\doibase 10.1016/0003-4916(59)90002-8} {\bibfield
  {journal} {\bibinfo  {journal} {Annals of Physics}\ }\textbf {\bibinfo
  {volume} {8}},\ \bibinfo {pages} {325 } (\bibinfo {year} {1959})}\BibitemShut
  {NoStop}%
\bibitem [{\citenamefont {Smith}\ and\ \citenamefont {Si}(2000)}]{Smith00}%
  \BibitemOpen
  \bibfield  {author} {\bibinfo {author} {\bibfnamefont {J.~L.}\ \bibnamefont
  {Smith}}\ and\ \bibinfo {author} {\bibfnamefont {Q.}~\bibnamefont {Si}},\
  }\href {\doibase 10.1103/PhysRevB.61.5184} {\bibfield  {journal} {\bibinfo
  {journal} {Phys. Rev. B}\ }\textbf {\bibinfo {volume} {61}},\ \bibinfo
  {pages} {5184} (\bibinfo {year} {2000})}\BibitemShut {NoStop}%
\bibitem [{\citenamefont {Chitra}\ and\ \citenamefont
  {Kotliar}(2001)}]{Chitra01}%
  \BibitemOpen
  \bibfield  {author} {\bibinfo {author} {\bibfnamefont {R.}~\bibnamefont
  {Chitra}}\ and\ \bibinfo {author} {\bibfnamefont {G.}~\bibnamefont
  {Kotliar}},\ }\href {\doibase 10.1103/PhysRevB.63.115110} {\bibfield
  {journal} {\bibinfo  {journal} {Phys. Rev. B}\ }\textbf {\bibinfo {volume}
  {63}},\ \bibinfo {pages} {115110} (\bibinfo {year} {2001})}\BibitemShut
  {NoStop}%
\bibitem [{\citenamefont {Rubtsov}\ \emph {et~al.}(2012)\citenamefont
  {Rubtsov}, \citenamefont {Katsnelson},\ and\ \citenamefont
  {Lichtenstein}}]{Rubtsov12}%
  \BibitemOpen
  \bibfield  {author} {\bibinfo {author} {\bibfnamefont {A.}~\bibnamefont
  {Rubtsov}}, \bibinfo {author} {\bibfnamefont {M.}~\bibnamefont {Katsnelson}},
  \ and\ \bibinfo {author} {\bibfnamefont {A.}~\bibnamefont {Lichtenstein}},\
  }\href {\doibase 10.1016/j.aop.2012.01.002} {\bibfield  {journal} {\bibinfo
  {journal} {Annals of Physics}\ }\textbf {\bibinfo {volume} {327}},\ \bibinfo
  {pages} {1320} (\bibinfo {year} {2012})}\BibitemShut {NoStop}%
\bibitem [{\citenamefont {Bulla}\ \emph {et~al.}(1998)\citenamefont {Bulla},
  \citenamefont {Hewson},\ and\ \citenamefont {Pruschke}}]{Bulla98}%
  \BibitemOpen
  \bibfield  {author} {\bibinfo {author} {\bibfnamefont {R.}~\bibnamefont
  {Bulla}}, \bibinfo {author} {\bibfnamefont {A.~C.}\ \bibnamefont {Hewson}}, \
  and\ \bibinfo {author} {\bibfnamefont {T.}~\bibnamefont {Pruschke}},\
  }\href@noop {} {\bibfield  {journal} {\bibinfo  {journal} {J. Phys. Condens.
  Matter}\ }\textbf {\bibinfo {volume} {10}},\ \bibinfo {pages} {8365}
  (\bibinfo {year} {1998})}\BibitemShut {NoStop}%
\bibitem [{\citenamefont {Toschi}\ \emph {et~al.}(2007)\citenamefont {Toschi},
  \citenamefont {Katanin},\ and\ \citenamefont {Held}}]{Toschi07}%
  \BibitemOpen
  \bibfield  {author} {\bibinfo {author} {\bibfnamefont {A.}~\bibnamefont
  {Toschi}}, \bibinfo {author} {\bibfnamefont {A.~A.}\ \bibnamefont {Katanin}},
  \ and\ \bibinfo {author} {\bibfnamefont {K.}~\bibnamefont {Held}},\ }\href
  {\doibase 10.1103/PhysRevB.75.045118} {\bibfield  {journal} {\bibinfo
  {journal} {Phys. Rev. B}\ }\textbf {\bibinfo {volume} {75}},\ \bibinfo {eid}
  {045118} (\bibinfo {year} {2007})}\BibitemShut {NoStop}%
\bibitem [{\citenamefont {Rubtsov}\ \emph {et~al.}(2008)\citenamefont
  {Rubtsov}, \citenamefont {Katsnelson},\ and\ \citenamefont
  {Lichtenstein}}]{Rubtsov08}%
  \BibitemOpen
  \bibfield  {author} {\bibinfo {author} {\bibfnamefont {A.~N.}\ \bibnamefont
  {Rubtsov}}, \bibinfo {author} {\bibfnamefont {M.~I.}\ \bibnamefont
  {Katsnelson}}, \ and\ \bibinfo {author} {\bibfnamefont {A.~I.}\ \bibnamefont
  {Lichtenstein}},\ }\href {\doibase 10.1103/PhysRevB.77.033101} {\bibfield
  {journal} {\bibinfo  {journal} {Phys. Rev. B}\ }\textbf {\bibinfo {volume}
  {77}},\ \bibinfo {eid} {033101} (\bibinfo {year} {2008})}\BibitemShut
  {NoStop}%
\bibitem [{\citenamefont {Hafermann}\ \emph {et~al.}(2009)\citenamefont
  {Hafermann}, \citenamefont {Li}, \citenamefont {Rubtsov}, \citenamefont
  {Katsnelson}, \citenamefont {Lichtenstein},\ and\ \citenamefont
  {Monien}}]{Hafermann09}%
  \BibitemOpen
  \bibfield  {author} {\bibinfo {author} {\bibfnamefont {H.}~\bibnamefont
  {Hafermann}}, \bibinfo {author} {\bibfnamefont {G.}~\bibnamefont {Li}},
  \bibinfo {author} {\bibfnamefont {A.~N.}\ \bibnamefont {Rubtsov}}, \bibinfo
  {author} {\bibfnamefont {M.~I.}\ \bibnamefont {Katsnelson}}, \bibinfo
  {author} {\bibfnamefont {A.~I.}\ \bibnamefont {Lichtenstein}}, \ and\
  \bibinfo {author} {\bibfnamefont {H.}~\bibnamefont {Monien}},\ }\href
  {\doibase 10.1103/PhysRevLett.102.206401} {\bibfield  {journal} {\bibinfo
  {journal} {Phys. Rev. Lett.}\ }\textbf {\bibinfo {volume} {102}},\ \bibinfo
  {pages} {206401} (\bibinfo {year} {2009})}\BibitemShut {NoStop}%
\end{thebibliography}%

\end{document}